\documentclass[lettersize,journal]{IEEEtran}

\usepackage{graphicx}
\usepackage{caption}
\usepackage{subcaption}

\usepackage{amsmath,amsfonts}
\usepackage{algorithmic}
\usepackage{algorithm}
\usepackage{array}

\usepackage{textcomp}

\usepackage{url}
\usepackage{amssymb}
\usepackage{verbatim}

\usepackage{svg}
\usepackage{stfloats} 

\usepackage{lettrine}

\usepackage{cite}
\begin{document}

\title{Movable Antenna-enabled RIS-aided Integrated
Sensing and Communication}

\author{\IEEEauthorblockN{Haisu Wu, Hong Ren,~\IEEEmembership{Member,~IEEE},  Cunhua Pan,~\IEEEmembership{Senior Member,~IEEE}, Yang Zhang}

\thanks{H. Wu is a bachelor's degree graduate from Shandong University, Weihai, China. (e-mail:  wuhaisu2020@gmail.com). He will join Southeast University as a Master student. H. Ren,  C. Pan and Y. Zhang are with National Mobile Communications Research Laboratory, Southeast University, Nanjing, China. (e-mail: {hren, cpan, 220230982}@seu.edu.cn).}
\thanks{\itshape{Corresponding authors:}  Hong Ren and Cunhua Pan.}
}

\markboth{}{}


\maketitle

\begin{abstract}
In this paper, we investigate a movable antenna (MA)-aided integrated sensing and communication (ISAC) system, where a reconfigurable intelligent surface (RIS) is employed to enhance  wireless communication and sensing performance in dead zones. Specifically, this paper aims to maximize the minimum beampattern gain at the RIS by jointly optimizing the covariance matrix at the base station (BS),  the reflecting coefficients at the RIS and the positions of the MAs,  subject to   signal-to-interference-plus-noise ratio (SINR) constraint for the users and  maximum transmit power at the BS. To tackle this non-convex optimization problem, we propose an alternating optimization (AO) algorithm and employ semidefinite relaxation (SDR), sequential rank-one constraint relaxation (SRCR) and  successive convex approximation (SCA) techniques. Numerical results indicate that the  MA-enabled RIS-aided ISAC system outperforms conventional fixed position antenna (FPA) and RIS-aided systems. In addition, the application of MAs can reconfigure  geometric properties of the antenna array and enhance channel gain in the ISAC system.

\end{abstract}


\section{Introduction}
\IEEEPARstart{F}{uture} sixth generation (6G) communication is anticipated to  support applications such as Massive Internet of Things (Massive-IoT), industrial automation (IA), and virtual reality (VR), which demand higher sensing precision and lower communication latency. These demands for 6G technologies call for a paradigm shift in the existing wireless communication networks. In particular, integrated sensing and communication (ISAC)  is regarded as one of the key transformative technologies for future 6G networks\cite{liu2018toward,liu2022integrated}. The objective of ISAC is to consolidate the sensing and communication functions onto a unified platform which  allows for the sharing of resources, hardware facilities, and signal-processing modules, thereby enhancing both spectrum and hardware efficiency. In addition,  millimeter-wave (mmW)/Terahertz (THz) signals \cite{akyildiz2022terahertz} and ultra-massive multiple-input
multiple-output (MIMO) \cite{faisal2020ultramassive} are expected to be utilized in the future communication system, implying that future wireless communication and radar sensing systems will become similar in hardware, which further enables the sharing of hardware resources.

Based on the aforementioned background, dual-functional radar and communication (DFRC) is proposed as a paradigm of ISAC systems, which performs both sensing and communication functions with shared frequency spectrum on an identical hardware platform \cite{hassanien2019dual,zheng2019radar}. Compared with another framework of ISAC, namely radar and communication coexistence (RCC), the integration of DFRC can reduce hardware costs and simplify design complexity. Currently, DFRC has received increasing attention, and most of the related investigations considered the beamforming design at the DFRC base station (BS). For instance, the authors of \cite{liu2018toward} initially considered the beampattern design in a DFRC system. Specifically, the spatial beamforming waveform at the DFRC BS is optimized to minimize the power of the interference for downlink communication users, subject to several design criteria of the radar.  In addition, the authors of \cite{Xia2020joint} investigated the beamforming design of a co-located MIMO system with a monostatic DFRC radar. Specifically, the authors proposed the simultaneous transmission of radar waveforms and communication symbols, which can form multiple beams and
better exploit the degrees of freedom (DoFs) in the MIMO system. Based on the aforementioned works, the authors of  \cite{hua2023optimal} considered the transmit beamforming in a downlink ISAC system to enhance the radar sensing performance while guaranteeing the user SINR constraint. In particular, the authors investigated the impact of whether the  receivers could cancel the interference from dedicated radar signals on sensing performance. Nevertheless, although mmW/THz signals can provide extensive bandwidth which is capable of enhancing the communication and sensing performance, the high frequency will lead to increased susceptibility of the transmit signal to be obstructed. In scenarios where the line-of-sight (LoS) links between the DFRC BS and the targets are obstructed, the performance of the ISAC system will experience substantial degradation.

Reconfigurable intelligent surface (RIS) is considered as a promising technology to address the aforementioned challenges \cite{zhang2020prospective,pan2021reconfigurable,pan2022overview}. An RIS is a meta-surface consisting of passive reflecting elements, each of which is capable of  independently altering the phases of incident signals. Therefore, the RIS can intelligently reconstruct the wireless channels, thereby strengthening the quality of service (QoS) of the legitimate users. In terms of the sensing function, RIS can also establish virtual LoS links in dead zones of the DFRC BS, significantly improving sensing capability in the ISAC systems \cite{song2022joint,sankar2023beamforming,wang2021joint,yu2023active}. For instance, the authors of \cite{song2022joint}  jointly optimized the transmit beamforming and the phase shifts of the RIS to maximize the minimum beampattern gain  towards the desired directions in the non-line-of-sight  (NLoS) area of the BS,  subject to communication SINR constraint of a single user. Furthermore, the authors of \cite{sankar2023beamforming} proposed the novel scheme of using two dedicated RISs for enhancing communication and  sensing tasks respectively. They also demonstrated that the sensing performance can be greatly enhanced when  targets are located at the NLoS area of the BS. 

Nonetheless, in the aforementioned ISAC systems, the MIMO architectures generally utilized fixed position antenna (FPA) arrays, which restrict the full utilization of the available DoFs within the continuous spatial domain, thereby hindering optimal spatial diversity and multiplexing performance in communication and sensing tasks. Furthermore, the unchangeable geometric configurations of conventional FPA arrays can lead to certain array gain loss during radar beamforming tasks. 

Recently, the novel concept of movable antenna (MA)  has been proposed as an innovative solution to overcome the inherent limitations in FPA systems \cite{wong2021FAS,ma2023mimo,zhu2023modeling}.  By facilitating a driver component or similar mechanisms, the positions of MAs can be adjusted dynamically within a designated spatial area, thereby reconstructing channel conditions to boost communication performance, or reconfiguring the geometric properties of the MIMO arrays to enhance sensing capability. Channel modeling for MA-aided communication systems was  explored in \cite{zhu2023modeling}, where the authors introduced a field-response channel model. They conducted a comparative performance analysis of  MA-aided and FPA systems across deterministic and statistical channels, highlighting the substantial advantages in enhancing received power and reducing outage probability of the  MA-aided system. In \cite{ma2023compressed}, \cite{zhang2024deep}, the research explored channel estimation for MA-aided communication systems by using compressed sensing and deep learning method, respectively.  Based on perfect channel state information (CSI), extensive research has shown that MA-aided systems provide substantial improvements over traditional FPA systems in wireless communication \cite{zhu2023multiuser,hu2024secure,tang2024secure,hu2024fluid,zheng2024fas}. For example,  MA-aided multi-terminal uplink transmission system with an FPA BS  was investigated  in \cite{zhu2023multiuser}, where a power minimization problem  was formulated subject to minimum-achievable-rate of terminals, and a gradient descent-based iterative method  was proposed to optimize the  positions of  MAs.

Nevertheless, the potential of MA  in the field of wireless sensing has not yet been fully explored, and only a few existing studies considered wireless sensing without communication \cite{ma2024multi},\cite{ma2024movable}. The authors of \cite{ma2024multi} investigated the enhanced multi-beam forming with a linear MA array by optimizing antenna positions to exploit new DoFs. In addition, in \cite{ma2024movable}, the authors delved into both one-dimensional (1D) and two-dimensional (2D) MA arrays and demonstrated that the MA arrays can greatly improve sensing capacity of the system over their FPA counterparts both analytically and numerically. Notably, numerical simulations revealed lower correlation of the steering vectors for 1D/2D MA arrays, which can enhance wireless sensing performance. Furthermore, the authors of \cite{lyu2024flexible}  introduced linear 1D MA into ISAC system, and then considered relatively simple settings where the transmit beamforming at the BS and the positions of the MAs were jointly optimized to maximize the communication rate and sensing mutual information. However, these studies all considered the LoS links between the BS and the targets. In contrast, if the targets may locate in the NLoS area of the BS, it is reasonable to apply an RIS to enhance the sensing performance of the MA-aided system.

Against the above background, this paper considers an ISAC  system which is assisted by a 2D MA array and an RIS for effective wireless sensing and communication in dead zones.  The main contributions of our paper are summarized as follows:
\begin{itemize}
  \item We consider an MA and RIS-aided ISAC system, where a DFRC BS equipped with a 2D MA array aims to sense the targets located in multiple angles and serve the communication users. In particular, the angles of interest are located in the NLoS area of the DFRC BS, thus an RIS is deployed to create a virtual LoS link to enhance the  wireless sensing and communication performance. Then, we formulate the minimum beampattern gain maximization problem subject to  communication SINR constraints.

  \item To address this highly non-convex optimization problem, we propose an AO-based algorithm which incorporates several  techniques: semidefinite relaxation (SDR),  sequential rank-one constraint relaxation (SRCR) and successive convex approximation (SCA). Specifically, we optimize the transmit beamforming at the BS by using SDR and  the  phase shifts at the RIS by using the SRCR algorithm to avoid potential non-convergence issues of SDR. For the tightly coupled  positions of MAs, we iteratively optimize individual antenna positions by using the SCA algorithm while fixing the positions of other antennas to obtain a suboptimal solution.

  \item Simulation results confirm the performance improvement of deploying MAs in RIS-aided ISAC systems compared to FPAs. In addition, the application of MAs can reduce the similarity of user channels in ISAC systems, thereby effectively
mitigating multi-user interference. Furthermore, antenna displacement can alleviate multipath effects caused by channel fading and enhance channel gain.
\end{itemize}

\textit{Notations:} Boldface lower case and upper case letters denote vectors and matrices, respectively. $(\cdot)^{*}$, $(\cdot)^T$ and $(\cdot)^H$ denote the conjugate, transpose and conjugate transpose (Hermitian), respectively. $\mathbb{E}[\cdot]$ denotes the expectation operation. $\|\boldsymbol{a}\|_2$ denotes the 2-norm of
vector $\boldsymbol{a}$. The set of $P\times Q$  complex and real matrices is denoted by $\mathbb{C}^{P\times Q}$ and $\mathbb{R}^{P\times Q}$, respectively. We use $\left[\boldsymbol{A}\right]_{p,q}$ to denote the entry of matrix $\boldsymbol{A}$ in its $p$-th row and $q$-th column. $\Re\{\boldsymbol{a}\}$ denotes the real part of vector $\boldsymbol{a}$. $\operatorname{diag}\left(\boldsymbol{a}\right)$ denotes a square diagonal matrix with the elements of vector $\boldsymbol{a}$ as its diagonal elements, while $\operatorname{diag}\left(\boldsymbol{A}\right)$ denotes a column vector consisting of the main diagonal elements of matrix $\boldsymbol{A}$. $\boldsymbol{A}\succeq0$ indicates that $\boldsymbol{A}$ is a positive semidefinite matrix. $\mathcal{CN}(0,\sigma^2)$ denotes the circularly symmetric complex Gaussian (CSCG) distribution with zero mean and covariance  $\sigma^2$.  $\operatorname{rank}\left(\boldsymbol{A}\right)$ denotes the rank of matrix $\boldsymbol{A}$. The amplitude and phase of complex value $a$ are denoted by $|a|$ and $\angle a$, respectively.

\section{System model and problem formulation}

We consider an MA and RIS-aided ISAC system as depicted in Fig. 1, where a DFRC BS equipped with a uniform linear array (ULA) of $M$ antennas serves $K$ single-antenna users and detects targets located in the NLoS area of the DFRC BS.  In addition,   an RIS with $N$ reflecting elements is deployed in the system to enhance the communication and sensing performance, and it is assumed that the targets are located at $L$ interested angles towards the RIS.

\begin{figure}
    \centering
    \includegraphics[width=1\linewidth]{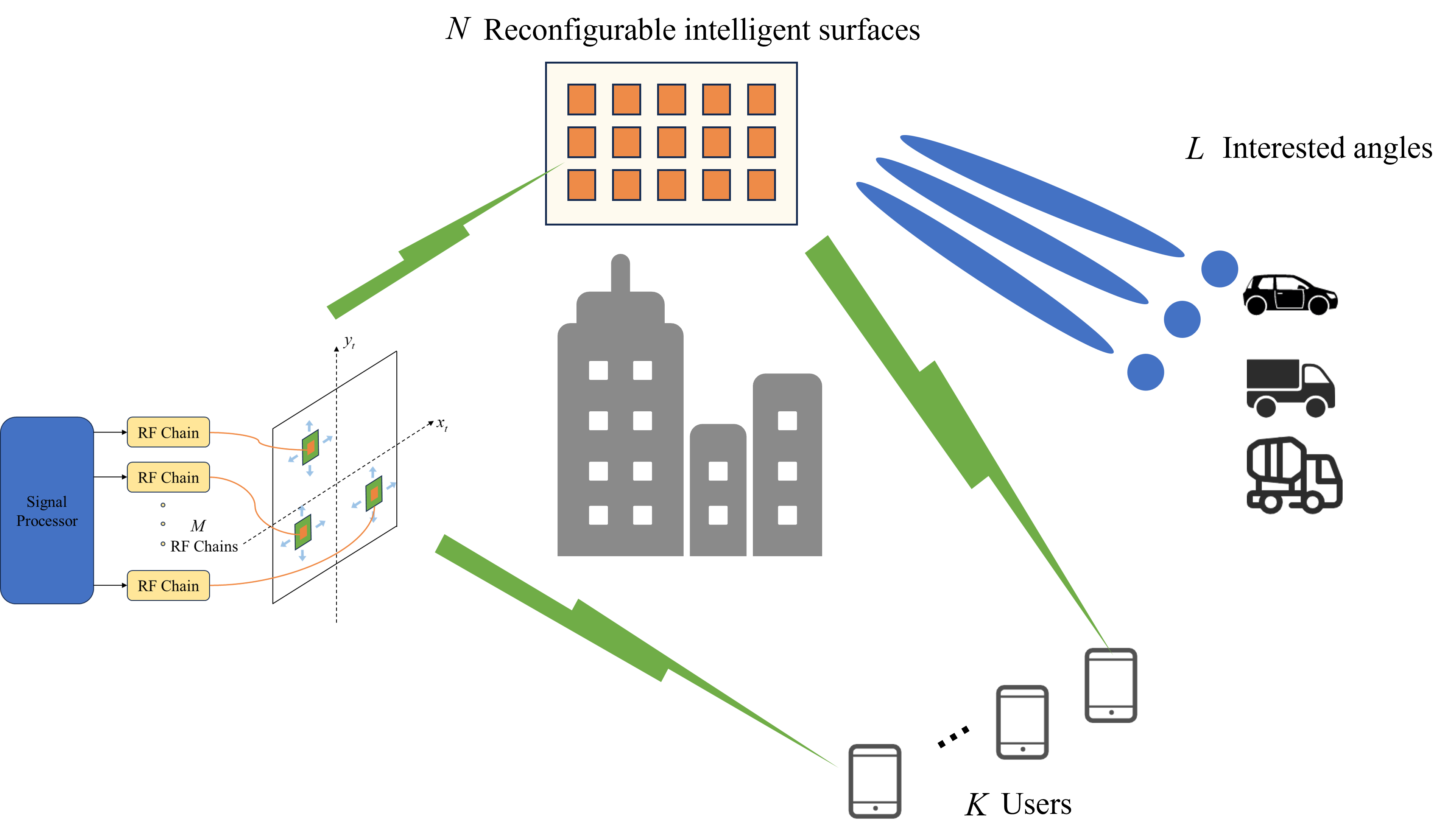}
    \caption{ MA and RIS-aided ISAC system.}
    \label{fig:system model}
\end{figure}

\subsection{Signal Model}

The transmit signal of the DFRC BS is represented as
\begin{equation}
\mathbf{x} = \mathbf{W}_{r} \mathbf{s}+\mathbf{W}_{c}\mathbf{c} =\left [ {\bf{W}}_{r},  {\bf{W}}_{c}\right ] \left[
\mathbf{s}^{T},  \mathbf{c}^{T}
\right]^{T}=\mathbf{W} \hat{\mathbf{x}},
\end{equation}

\noindent where $\mathbf{s} \in \mathbb{C}^{M \times 1}$ is the radar signal, and $\mathbf{c} \in \mathbb{C}^{K \times 1}$ denotes the transmit symbols to the  $K$ users. In addition, $\mathbf{W}_{r}=\left[\mathbf{w}_{r, 1}, \mathbf{w}_{r, 2}, \ldots, \mathbf{w}_{r, M}\right] \in \mathbb{C}^{M \times M}$, $\mathbf{W}_{c}=\left[\mathbf{w}_{c, 1}, \mathbf{w}_{c, 2}, \ldots, \mathbf{w}_{c, K}\right] \in \mathbb{C}^{M \times K}$  denote the beamforming matrices for radar and communication, respectively.
$\mathbf{W} = [\mathbf{W}_r, \mathbf{W}_c]\in \mathbb{C}^{M \times(M+K)}$ represents the equivalent DFRC transmit beamforming matrix. It is assumed that the radar signals are generated by pseudo-random coding, which satisfies $\mathbb{E}[\mathbf{s}]=\mathbf{0}$ and $\mathbb{E}\left[\mathbf{s s}^{H}\right]=\mathbf{I}_{M}$. The communication signal $\bf c$ is assumed to follow $\mathcal{C N}\left(\mathbf{0}, \mathbf{I}_{K}\right)$, and the radar and communication signals are uncorrelated. Therefore, the covariance matrix of the transmit signal can be expressed as
\begin{equation}
\mathbf{R}=\mathbb{E}\left[\mathbf{x x}^{H}\right]=\mathbf{W} \mathbf{W}^{H}=\mathbf{W}_{r} \mathbf{W}_{r}^{H}+\sum_{k=1}^{K} \mathbf{R}_{k},
\end{equation}
where the rank-one matrix $\mathbf{R}_{k}$ is introduced by $\mathbf{R}_{k} \triangleq \mathbf{w}_{c, k} \mathbf{w}_{c, k}^{H}$.

\subsection{Channel Model}

\begin{figure}
    \centering
    \includegraphics[width=0.75\linewidth,keepaspectratio]{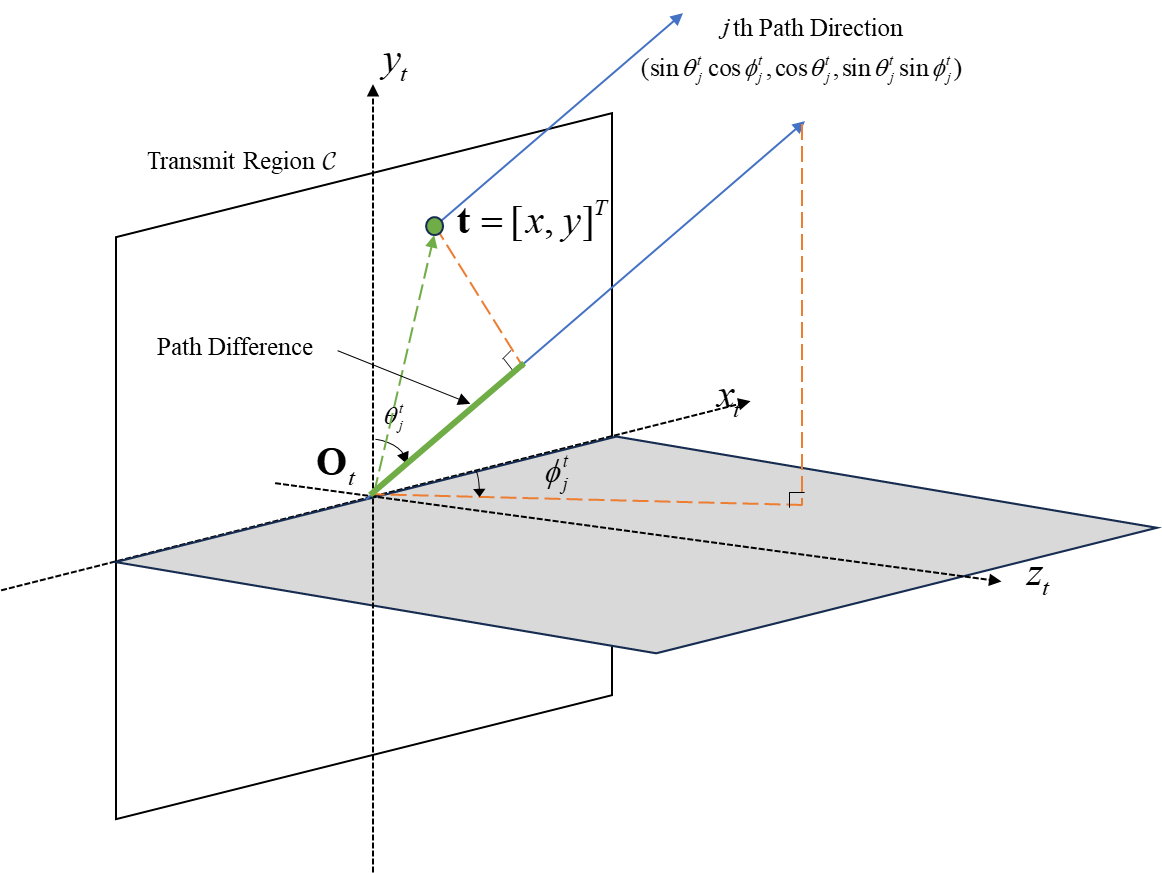}
    \caption{Illustration of the coordinates and spatial angles in transmit region.}
    \label{fig:lauch area sketch}
\end{figure}

This paper utilizes the planar far-field response model  in \cite{zhu2023modeling}, where each  transmit/receive path of the same channel has the same angle of departure (AoD), angle of arrival (AoA), and path response coefficient, but different signal phases. Specifically,  let us define $\tilde{\boldsymbol{t}} = [\boldsymbol{t}_1, \boldsymbol{t}_2, \ldots, \boldsymbol{t}_M] \in \mathbb{R}^{2 \times M}$ as the positions of MAs at the BS, where $\boldsymbol{t}_m = [x_m, y_m]^T \in \mathcal{C}, 1 \leq m \leq M$ represents the coordinates of the $m$-th MA, and $\mathcal{C}$ is the transmit region of the MAs.

As shown in  Fig. \ref{fig:lauch area sketch}, the elevation and azimuth angles of the transmit area and the receive area are denoted as $\theta_j^t \in [0, \pi], \phi_j^t \in [0, \pi], 1 \leq j \leq L_t$ and $\theta_i^r \in [0, \pi], \phi_i^r \in [0, \pi], 1 \leq i \leq L_r$, respectively. Here, $L_t$ is the number of transmit paths from the BS to the RIS, and $L_r$ is the number of receive paths at the RIS, respectively. In addition, the number of transmit paths for the direct channel from the BS to the $k$-th user is defined as $L^k_t, 1 \leq k \leq K$. For the MA-aided system, the channel matrix is determined by the signal propagation environment and the MAs' positions. According to \cite{zhu2023modeling}, the field response vector (FRV) of the MA  at position $\boldsymbol{t} = [x, y]^T$ in the BS-RIS link is given by
\begin{equation}
\boldsymbol{g}(\boldsymbol{t})=\left[e^{\mathrm{j} \frac{2 \pi}{\lambda} \rho_{t}^{1}(\boldsymbol{t})}, e^{\mathrm{j} \frac{2 \pi}{\lambda} \rho_{t}^{2}(\boldsymbol{t})}, \ldots, e^{\mathrm{j} \frac{2 \pi}{\lambda} \rho_{t}^{L_{t}}(\boldsymbol{t})}\right]^{T} \in \mathbb{C}^{L_{t}\times1},
\end{equation}
where $\lambda$ represents the wavelength and $\rho_t^j(\boldsymbol{t}) = x \sin \theta_j^t \cos \phi_j^t + y \cos \theta_j^t$ represents the difference of the signal propagation distance for the  $j$-th transmit path between position $\boldsymbol{t}$ and reference origin. Therefore, the field response matrix (FRM) of the BS-RIS link for all $M$ transmit MAs is given by
\begin{equation}
\boldsymbol{G}(\tilde{\boldsymbol{t}}) \triangleq\left[\boldsymbol{g}\left(\boldsymbol{t}_{1}\right), \boldsymbol{g}\left(\boldsymbol{t}_{2}\right), \ldots, \boldsymbol{g}\left(\boldsymbol{t}_{M}\right)\right] \in \mathbb{C}^{L_{t} \times M}.
\end{equation}
Similarly, the FRM of the BS-User link for all $M$  MAs is:
\begin{equation}
\boldsymbol{G}_k(\tilde{\boldsymbol{t}}) \triangleq \left[ \boldsymbol{g}_k(\boldsymbol{t}_1), \boldsymbol{g}_k(\boldsymbol{t}_2), \ldots, \boldsymbol{g}_k(\boldsymbol{t}_M) \right] \in \mathbb{C}^{L_t^k \times M}, 1 \leq k \leq K.
\end{equation}
In addition, it can be derived that  the FRV for a single reflecting element at the RIS is
\begin{equation}
\boldsymbol{f}\left(\boldsymbol{r}\right)=\left[\mathrm{e}^{\mathrm{j} \frac{2 \pi}{\lambda} \rho_{r}^{1}\left(\boldsymbol{r}\right)}, \mathrm{e}^{\mathrm{j} \frac{2 \pi}{\lambda} \rho_{r}^{2}\left(\boldsymbol{r}\right)},\cdots, \mathrm{e}^{\mathrm{j} \frac{2 \pi}{\lambda} \rho_{r}^{L_r}\left(\boldsymbol{r}\right)}\right]^{T}\in \mathbb{C}^{L_{r}\times 1},
\end{equation}

\noindent where $\rho_{r}^{i}(\boldsymbol{r})=x\sin\theta_{i}^{r}\cos\phi_{i}^{r}+y\cos\theta_{i}^{r}$ is the difference of the signal propagation distance for the $i$-th receive path
between the element position $\boldsymbol{r}$ and the origin of the RIS. Hence, the FRM at the RIS is represented as
\begin{equation}
\boldsymbol{F}(\boldsymbol r)=\left[\boldsymbol{f}\left(\boldsymbol{r}_{1}\right), \boldsymbol{f}\left(\boldsymbol{r}_{2}\right), \cdots, \boldsymbol{f}\left(\boldsymbol{r}_{N}\right)\right] \in \mathbb{C}^{L_r \times N}.
\end{equation}

\noindent Then, let us denote the path response matrix (PRM) $\mathbf{\Sigma} \in \mathbb{C}^{L_r \times L_t}$ and $\mathbf{\Sigma}_k \in \mathbb{C}^{L^k_r \times L^k_t}, 1 \leq k \leq K$ as the responses of all transmit and receive paths from the BS to the RIS and from the BS to the $k$-th user, respectively. Therefore, the channel matrix from the BS to the RIS can be expressed as
\begin{equation}
\mathbf{H}(\tilde{\boldsymbol{t}})=\boldsymbol{F}(\boldsymbol{r})^{H} \boldsymbol{\Sigma} \boldsymbol{G}(\tilde{\boldsymbol{t}}).
\end{equation}
Similarly, the channel from the BS to the $k$-th user can be given by 
\begin{equation}
\mathbf{h}_{1, k}^H(\tilde{\boldsymbol{t}}) = \boldsymbol{1}^H\boldsymbol{\Sigma}_k\boldsymbol{G}_k(\tilde{\boldsymbol{t}})\in \mathbb{C}^{1 \times M}.
\end{equation}
Furthermore, the positions of RIS elements and antennas of the users  are fixed, and the channel matrix from the RIS to the $k$-th user can be represented as  $\mathbf{h}_{2, k}^H \in \mathbb{C}^{1 \times N }$. We also assume that the CSI of the channels is perfectly known at the DFRC BS with the application of  MA-related channel estimation methods \cite{ma2023compressed},\cite{zhang2024deep}.

\subsection{Communication and Radar Sensing Metrics}

By combining the signals from both BS-User and BS-RIS-User links, the received signal at the $k$-th user is
\begin{equation}
y_k=\left(\mathbf{h}^H_{2, k}\boldsymbol{\Phi} \mathbf{H}(\tilde{\boldsymbol{t}})+\mathbf{h}_{1, k}^H(\tilde{\boldsymbol{t}})\right)\mathbf{x}+n_k,
\end{equation}
where $n_k\sim\mathcal{CN}(0,\sigma^2_k)$ denotes additive white Gaussian noise (AWGN) at the $k$-th user receiver. Thus, the SINR for the $k$-th user is
\begin{equation}
\begin{aligned}
\mathrm{SINR}_{k} & =\frac{ \mathbf{h}_{k}^{H} \mathbf{w}_{k} \mathbf{w}_{k}^{H} \mathbf{h}_{k}}{\sum_{1 \leq i \leq M+K, i \neq k} \mathbf{h}_{k}^{H} \mathbf{w}_{i} \mathbf{w}_{i}^{H} \mathbf{h}_{k}+\sigma^{2}_k} \\
& =\frac{\mathbf{h}_{k}^{H} \mathbf{R}_{k}\mathbf{h}_{k}}{\sum_{1 \leq i \leq M+K, i \neq k}\mathbf{h}_{k}^H \mathbf{R}_{i} \mathbf{h}_{k}+\sigma^{2}_k} \\
& =\frac{\mathbf{h}_{k}^{H} \mathbf{R}_{k}\mathbf{h}_{k}}{{\mathbf{h}_{k}^{H}\left(\mathbf{R}-\mathbf{R}_{k}\right) \mathbf{h}_{k}+\sigma^{2}_k}},
\end{aligned}
\end{equation}

\noindent where $\mathbf{h}_{k}^{H}(\tilde{\boldsymbol{t}}) \triangleq \mathbf{h}_{2, k}^{H} \boldsymbol{\Phi} \mathbf{H}(\tilde{\boldsymbol{t}})+\mathbf{h}_{1, k}^{H}(\tilde{\boldsymbol{t}})$ represents the equivalent channel between the BS and the $k$-th user.

Next, we consider the perception of potential targets in the NLoS area of BS. In this case,  we assume that the virtual LoS channel created by the RIS are much stronger than the NLoS one. Hence, the effect of NLoS channel between the BS and sensing targets can be neglected. In addition, it is assumed that the location of the  RIS is well designed with few obstacles, and we adopt the beam direction gain of the RIS at the desired sensing angles as the sensing performance metrics. Let $d_{\text{RIS}}$ represent the spacing between consecutive reflecting elements at the RIS, and the steering vector at the RIS is given by
\begin{equation}
\boldsymbol{a}(\theta)=\left[1, e^{j \frac{2 \pi d_{\mathrm{RIS}}}{\lambda} \sin \theta}, \ldots, e^{j \frac{2 \pi(N-1) d_{\mathrm{RIS}}}{\lambda} \sin \theta}\right]^{T},
\end{equation}
where $\theta$ denotes the AoD of the target towards the RIS. We also consider that both the signal symbol $\mathbf{c}$ and the radar symbol $\mathbf{s}$ can be used for sensing\cite{Xia2020joint}. Therefore, the beampattern gain at the RIS corresponding to angle $\theta$ is
\begin{equation}
    \begin{aligned}
        \mathcal{P}(\theta) & =\mathbb{E}\left(\left|\boldsymbol{a}(\theta)^{H} \boldsymbol{\Phi} \mathbf{H}(\tilde{\boldsymbol{t}})\left(\mathbf{W}_{r} \mathbf{s}+\mathbf{W}_{c} \mathbf{c}\right)\right|^{2}\right) \\
        & =\boldsymbol{a}(\theta)^{H} \boldsymbol{\Phi} \mathbf{H}(\tilde{\boldsymbol{t}})\mathbf{R} \mathbf{H}(\tilde{\boldsymbol{t}})^{H} \boldsymbol{\Phi}^{H} \boldsymbol{a}(\theta) .
        \end{aligned}\label{eq:beampattern}
\end{equation}

We are interested in   $L$ specific angles $\{\theta_1,\cdots,\theta_L\}$, and we define $\mathcal{L} \triangleq\{1, \cdots, L\}$ as the set of sensing angles of interest. 
\subsection{Problem Formulation}
In this paper, we aim to maximize the minimum beampattern gain in the directions of the $L$ interested sensing angles by jointly optimizing  the covariance matrix $\bf R$ of the transmit beamformer  at the BS, the phase shifts $\mathbf{\Phi}$ at the RIS, and the positions $\tilde{\boldsymbol{t}}$ of the  MAs.  Accordingly, the problem is formulated as 
\begin{subequations}
\begin{align}
 \max _{ \mathbf{R}, \boldsymbol{\Phi},\tilde{\boldsymbol{t}}}   \quad & \min _{l \in \mathcal{L}} \boldsymbol{a}(\theta)^{H} \boldsymbol{\Phi} \mathbf{H}(\tilde{\boldsymbol{t}})\mathbf{R} \mathbf{H}(\tilde{\boldsymbol{t}})^{H} \boldsymbol{\Phi}^{H} \boldsymbol{a}(\theta_l) \\
\text{s.t.}\quad  & \mathrm{SINR}_{k} \geq \Gamma,  \forall k,\label{eq:SINR}\\
& \operatorname{tr}\left(\mathbf{R}\right) \leq P_{0}, \\
& \mathbf{R} \succeq 0, \\
& \operatorname{rank}(\mathbf{R}_k) = 1 \\
& \boldsymbol{\Phi}=\operatorname{diag}\left(e^{j \phi_{1}}, \ldots, e^{j \phi_{N}}\right),\\
& \left\|\boldsymbol{t}_{k}-\boldsymbol{t}_{q}\right\|_{2} \geq D, \quad  k \neq q,\\
& \tilde{\boldsymbol{t}} \in \mathcal{C},
\end{align}\label{eq:op_problem}%
\end{subequations}where $P_0$ denotes  the maximum transmit power  at the BS, $\Gamma$ denotes the minimum SINR threshold of the users, and $D$ represents the minimum distance  between  MAs to prevent coupling effects. 

It is challenging to solve Problem (\ref{eq:op_problem})  due to the intricately coupled variables $\mathbf{R}$, $\mathbf\Phi$, $\tilde{\boldsymbol{t}}$. Furthermore, the objective function of Problem (\ref{eq:op_problem}) exhibits non-smooth characteristics, and the channel vectors are highly non-convex with respect to  MA positions $\tilde{\boldsymbol{t}}$, which increases the difficulty of solving Problem (\ref{eq:op_problem}). 

\section{PROPOSED AO-BASED ALGORITHM}
In this section, we decouple the original problem into three subproblems and optimize them alternately. Specifically, we employ the SDR algorithm to optimize the transmit covariance matrix $\mathbf{R}$, and utilize the SRCR algorithm to optimize the reflecting coefficient $\mathbf \Phi$. Subsequently, the positions of   MAs, denoted as $\tilde{\boldsymbol{t}}$, are optimized by using the SCA technique. 
\subsection{Transmit Beamforming Optimization at the BS}
In this subsection, we optimize the covariance matrix $\bf R$ at the DFRC BS with fixed $\bf \Phi$ and $\tilde{\boldsymbol{t}}$. To handle the  requirements of  SINR (\ref{eq:SINR}), we can transform the SINR constraint of $k$-th user into the following form
\begin{equation}
\left(1+\Gamma^{-1}\right) \operatorname{tr}(\mathbf R_k \mathbf H_k)  \geq \operatorname{tr}(\mathbf R\mathbf H_k)+\sigma_k^{2}, \forall k ,
\end{equation}
\noindent where $\mathbf H_k=\mathbf{h}_{k}\mathbf{h}_{k}^H$.
As a result, Problem (\ref{eq:op_problem}) can be reformulated as
\begin{subequations}
\begin{align}
\max _{ \mathbf{R}} \quad & \min _{l \in \mathcal{L}} \boldsymbol{a}(\theta)^{H} \boldsymbol{\Phi} \mathbf{H}(\tilde{\boldsymbol{t}})\mathbf{R} \mathbf{H}(\tilde{\boldsymbol{t}})^{H} \boldsymbol{\Phi}^{H} \boldsymbol{a}(\theta_l)\\
\text{s.t.} \quad & \left(1+\Gamma^{-1}\right) \operatorname{tr}(\mathbf R_k \mathbf H_k)  \geq \operatorname{tr}(\mathbf R\mathbf H_k)+\sigma_k^{2},\forall k ,\label{eq:BS_beamforming_sinr}\\
& \operatorname{tr}\left(\mathbf{R}\right) \leq P_{0},\label{eq:BS_beamforming_power} \\
& \mathbf{R} \succeq 0, \label{eq:BS_beamforming_semi}\\
& \operatorname{rank}(\mathbf{R}_k) = 1.\label{eq:BS_beamforming_rank}
\end{align}\label{eq:BS_beamforming}%
\end{subequations}
Owing to the  rank-one constraint (\ref{eq:BS_beamforming_rank}), Problem (\ref{eq:BS_beamforming}) is still non-convex. To handle this problem, we utilize the key idea of the SDR method. By omitting the rank-one constraints, Problem (\ref{eq:BS_beamforming}) can be recast as
\begin{subequations}
\begin{align}
\max _{ \mathbf{R},\{\mathbf{R}_k\}}   \quad & \min _{l \in \mathcal{L}} \operatorname{tr}(\mathbf A(\theta_l)\mathbf{R})  \\
\text{s.t.} \quad & \left(1+\Gamma^{-1}\right) \operatorname{tr}(\mathbf R_k \mathbf H_k)\nonumber\label{eq:BS_beamforming_SDR_sinr} \\
& \geq \operatorname{tr}(\mathbf R\mathbf H_k)+\sigma_k^{2}, \forall k, \\
& \operatorname{tr}\left(\mathbf{R}\right) \leq P_{0}, \\
& \mathbf{R} \succeq 0,\label{eq:eq:BS_beamforming_SDR_semi} \\
&\mathbf{R}-\sum_{k=1}^{K} \mathbf{R}_{k} \succeq 0,
\end{align}\label{eq:BS_beamforming_SDR}
\end{subequations}

\noindent where $\mathbf A(\theta_l)=\mathbf{H}^{H} \boldsymbol{\Phi}^{H} \boldsymbol{a}(\theta_l)\boldsymbol{a}(\theta)^{H} \boldsymbol{\Phi} \mathbf{H}$ is  independent of $\mathbf{R}$. Problem (\ref{eq:BS_beamforming_SDR}) is a convex SDP problem where all the constraints are either linear or semidefinite and the global optimal point can be obtained in polynomial time by using the convex optimization toolbox CVX\cite{boyd2004convex}. Denoting $\hat{\mathbf{R}},\{\hat{\mathbf{R}}_k\}$ as a feasible solution to Problem (\ref{eq:BS_beamforming_SDR}), we introduce the following lemma to construct the corresponding rank-one solution $\bar{\mathbf{R}},\{\bar{\mathbf{R}}_k\}$ to Problem (\ref{eq:BS_beamforming}).

\itshape \textbf{Lemma:}  \upshape There exists a global optimum $\bar{\mathbf{R}},\{\bar{\mathbf{R}}_k\}$  for Problem (\ref{eq:BS_beamforming}),  satisfying
\begin{equation}
	 \operatorname{rank}(\bar{\mathbf{R}}_k) = 1,\forall k.
\end{equation}

\itshape \textbf{Proof:}  \upshape Please refer to\cite{yu2023active}. \hfill $\blacksquare$

\subsection{Reflective Beamforming Optimization}
In this subsection,  we optimize the phase shift  $\mathbf \Phi$  of the RIS, with fixed transmit covariance matrix $\mathbf R$  and fixed position $\tilde{\boldsymbol{t}}$ of $M$  MAs. The subproblem can be rewritten as 
\begin{subequations}
\begin{align}
\max _{  \boldsymbol{\Phi}}   \quad & \min _{l \in \mathcal{L}} \boldsymbol{a}(\theta)^{H} \boldsymbol{\Phi} \mathbf{H}(\tilde{\boldsymbol{t}})\mathbf{R} \mathbf{H}(\tilde{\boldsymbol{t}})^{H} \boldsymbol{\Phi}^{H} \boldsymbol{a}(\theta_l) \\
\text{s.t.}\quad  & \mathrm{SINR}_{k} \geq \Gamma,  \forall k,\label{eq:RIS_SINR}\\
& \boldsymbol{\Phi}=\operatorname{diag}\left(e^{j \phi_{1}}, \ldots, e^{j \phi_{N}}\right).
\end{align}\label{eq:RIS}%
\end{subequations}
\setlength{\parindent}{1em}
We first define $\boldsymbol{v}\triangleq \operatorname{vec}({\boldsymbol{\Phi}^{*}})=\left[e^{j \phi_{1}}, \ldots, e^{j \phi_{N}}\right]^{H}$ and introduce the following definition \begin{equation}
\mathbf{H}_l\triangleq \operatorname{diag}\left(\boldsymbol{\alpha}\left(\theta_{l}\right)^{H} \right)\mathbf{H}(\tilde{\boldsymbol{t}}) \mathbf{R} \mathbf{H}(\tilde{\boldsymbol{t}})^{H} \operatorname{diag}\left(\boldsymbol{\alpha}\left(\theta_{l}\right)\right),  \forall l \in \mathcal{L}.
\end{equation}
Then the beampattern gain towards angle $\theta_l$ is rewritten as \begin{equation}
\mathcal{P}(\theta_l)=\boldsymbol{v}^{H} \mathbf{H}_l \boldsymbol{v}.\label{eq:P}   
\end{equation}
Next, by letting $\boldsymbol G_k=\operatorname{diag}(\mathbf h_{1,k}^H) \mathbf{H}(\tilde{\boldsymbol{t}})$, the SINR constraint in (\ref{eq:SINR}) is reformulated as \begin{equation}
    \left(\boldsymbol{v}^{H} \boldsymbol{G}_k+\mathbf{H}_{2,k}^H\right) \widetilde{ \mathbf{R}}_k\left(\boldsymbol{G}_k^{H} \boldsymbol{v}+\mathbf{H}_{2,k}\right) \geq \sigma^{2},\forall k,\label{eq:RIS_SINR}
\end{equation}
where $\widetilde{\mathbf R}_k=(1+\Gamma^{-1})\mathbf R_k-\mathbf R$. In addition, (\ref{eq:RIS_SINR}) can be further recast as \begin{equation}  \bar{\boldsymbol{v}}^{H} \boldsymbol{W}_{k} \bar{\boldsymbol{v}} \geq \sigma_k^{2},\forall k,
\end{equation}
with \begin{equation}
\boldsymbol{W}_{k}=\left[\begin{array}{cc}
\boldsymbol{G}_k \widetilde{\mathbf{R}} _k \boldsymbol{G}_k^{H} & \boldsymbol{G}_k\widetilde{\mathbf{R}} _k \mathbf{H}_{2,k}\\
\mathbf{H}_{2,k}^H \widetilde{\mathbf{R}} _k \boldsymbol{G}_k^{H} & \mathbf{H}_{2,k}^H\widetilde{\mathbf{R}} _k\mathbf{H}_{2,k}
\end{array}\right],\bar{\boldsymbol{v}}=\left[\begin{array}{l}
\boldsymbol{v} \\
1
\end{array}\right].\label{eq:bar_v}
\end{equation}
For further manipulations, let us define\begin{equation}
\bar{\mathbf{H}}_l=\left[\begin{array}{cc}
\mathbf{H}_{l}& \mathbf{0}_{N \times 1} \\
\mathbf{0}_{1 \times N} & 0
\end{array}\right],\label{eq:bar_H_l}
\end{equation}
and by subsituting (\ref{eq:bar_v}), (\ref{eq:bar_H_l}) into (\ref{eq:P}), we have $\mathcal{P}(\theta)=\overline{\boldsymbol{v}}^{H} \bar{\mathbf{H}}_l\overline{\boldsymbol{v}}$. Hence, Problem (\ref{eq:RIS})  can be recast as
\begin{subequations}
\begin{align}
\max _{ \bar{\boldsymbol{v}}} \quad & \min _{l \in \mathcal{L}}\overline{\boldsymbol{v}}^{H} \bar{\mathbf{H}}_l\overline{\boldsymbol{v}} \\
\text{s.t.}\quad  & \bar{\boldsymbol{v}}^{H} \boldsymbol{W}_{k} \bar{\boldsymbol{v}} \geq \sigma^{2}_k,  \forall k,\\
& \left|\overline{\boldsymbol{v}}_{n}\right|=1, \forall n \in\{1, \ldots, N+1\},\label{eq:unit_modulus}
\end{align}\label{eq:RIS_unit_modulus}%
\end{subequations}
which is still  non-convex due to  the unit-modulus constraints  (\ref{eq:unit_modulus}). To address this issue, we define $\bar{\boldsymbol{V}}=\bar{\boldsymbol{v}} \bar{\boldsymbol{v}}^{H}$ and Problem (\ref{eq:RIS_unit_modulus}) is reformulated as 
\begin{subequations}
\begin{align}
 \max _{ \bar{\boldsymbol{v}}} \quad & \min _{l \in \mathcal{L}} \operatorname{tr}(\bar{\mathbf{H}}_l\bar{\boldsymbol{V}})  \\
\text{s.t.} \quad & \operatorname{tr}( \boldsymbol{W}_{k} \bar{\boldsymbol{V}}) \geq \sigma_k^{2},  \forall k, \\
& \bar{\boldsymbol{V}}_{n, n}=1, \forall n \in\{1, \ldots, N+1\},\\
&\bar{\boldsymbol{V}} \succeq0,\\
&\operatorname{rank}(\bar{\boldsymbol{V}})=1.\label{eq:rank1}
\end{align}\label{eq:RIS_SDP}%
\end{subequations}
The only non-convex constraint in (\ref{eq:RIS_SDP}) is the rank-one constraint (\ref{eq:rank1}), which can  be typically addressed by using the SDR and Gaussian randomization  technique. However, in the case of strict constraints, as in Problem (\ref{eq:RIS_SDP}), the approximation methods such as  Gaussian randomization algorithm  cannot guarantee the convergence of the overall algrithm\cite{wang2021joint}. Therefore, we adopt SRCR  to convert (\ref{eq:rank1}) equivalently to\cite{zuo2023exploi}
\begin{equation}
    \boldsymbol u_{\max}^{H}\left(\bar{\boldsymbol{V}}^{(t)}\right)\bar{\boldsymbol{V}}\boldsymbol u_{\max}\left(\bar{\boldsymbol{V}}^{(t)}\right)\geqslant w^{(t)}\mathrm{Tr}\left(\bar{\boldsymbol{V}}\right),
\end{equation}
where $\bar{\boldsymbol{V}}^{(t)}$ is a feasible solution obtained in the $t$-th iteration, $\boldsymbol u_{\max}\left(\bar{\boldsymbol{V}}^{(t)}\right)$ is the eigenvector correspoding to the maximum eigenvalue of $\bar{\boldsymbol{V}}^{(t)}$, $w^{(t)}$ denotes a relaxation parameter which will gradually approach 1. Finally, our optimization problem in the $t$-th iteration is given by 
\begin{subequations}
    \begin{align}
\max _{ \bar{\boldsymbol{V}}} \quad & \min _{l \in \mathcal{L}} \operatorname{tr}(\bar{\mathbf{H}}_l\bar{\boldsymbol{V}}) \\
\text{s.t.}\quad  & \operatorname{tr}( \boldsymbol{W}_{k} \bar{\boldsymbol{V}}) \geq \sigma^{2}_k,  \forall k, \\
& \bar{\boldsymbol{V}}_{n, n}=1, \forall n \in\{1, \ldots, N+1\},\\
&\bar{\boldsymbol{V}} \succeq 0,\\
& \boldsymbol u_{\max}^H\left(\bar{\boldsymbol{V}}^{(t)}\right)\bar{\boldsymbol{V}}\boldsymbol u_{\max}\left(\bar{\boldsymbol{V}}^{(t)}\right)\geqslant w^{(t)}\mathrm{Tr}\left(\bar{\boldsymbol{V}}\right).
\end{align}\label{eq:t_scrc}%
\end{subequations}
Problem (\ref{eq:t_scrc}) is an SDP problem and can be solved by the CVX tool \cite{boyd2004convex}. The procedure for
optimizing RIS reflecting coefficients is summarized in Algorithm 1.

\begin{algorithm}[!t]
	\caption{SRCR Algorithm for Optimizing RIS Coefficients}
	\label{alg1}
	\begin{algorithmic}[1]
		\STATE \emph{Input:} $\{\bar{\mathbf{H}}_l\}_{l=1}^L$, $\{\boldsymbol{W}_{k}\}_{k=1}^K$.
        \STATE Initialize $\{\boldsymbol{t}_m\}_{n=1}^M$, $w^{(0)}=0$, $\tau^{(0)}$.   \WHILE{$\mathrm{Tr}\left(\bar{\boldsymbol{V}}^{(t)}\right)/\lambda_{\max}\Big(\bar{\boldsymbol{V}}^{(t)}\Big)-1$ is above $\epsilon$}
            \IF{ Problem (\ref{eq:t_scrc}) is solvable with $\boldsymbol{V}^{(t)}$ and $w^{(t)}$}
            \STATE Obtain $\bar{\boldsymbol{V}}^{(t+1)}$ by solving Problem (\ref{eq:t_scrc}), $\tau^{(t+1)}$=$\tau^{(0)}$
            \ELSE 
            \STATE Update $\bar{\boldsymbol{V}}^{(t+1)}$=$\bar{\boldsymbol{V}}^{(t)}$;
            \STATE Update $\tau^{(t+1)}$=$\tau^{(t)}/2$;
            \ENDIF
            
            \STATE $w^{(t+1)}=\min\left(1,\frac{\lambda_{\max}\left(\bar{\boldsymbol{V}}^{(t+1)}\right)}{\mathrm{Tr}\left(\bar{\boldsymbol{V}}^{(t+1)}\right)}+\tau^{(t+1)}\right)$;
            \STATE $t=t+1$;
        \ENDWHILE
        \STATE Obtain $\boldsymbol{\Phi}$ by decompose $\bar{\boldsymbol{V}}$.
        \STATE \emph{Output:} $\boldsymbol{\Phi}$.
	\end{algorithmic}
\end{algorithm}
\subsection{Antenna Position Design}

In this subsection, we optimize the position $\tilde{\boldsymbol{t}}$ of $M$  MAs, with fixed transmit covariance matrix $\mathbf R$ and reflecting coefficient $\mathbf \Phi$. The optimization problem can be
further reformulated as
\begin{subequations}
\begin{align}
\max _{ \tilde{\boldsymbol{t}}} \quad & \min _{l \in \mathcal{L}} \boldsymbol{a}(\theta)^{H} \boldsymbol{\Phi} \mathbf{H}(\tilde{\boldsymbol{t}})\mathbf{R} \mathbf{H}(\tilde{\boldsymbol{t}})^{H} \boldsymbol{\Phi}^{H} \boldsymbol{a}(\theta_l) \\
\text{s.t.}\quad  & \mathrm{SINR}_{k} \geq \Gamma, \forall k,\\ 
& \left\|\boldsymbol{t}_{k}-\boldsymbol{t}_{q}\right\|_{2} \geq D,   k \neq q,\\
& \tilde{\boldsymbol{t}} \in \mathcal{C},
\end{align}\label{eq:t}%
\end{subequations}
which is a non-convex problem owing to the complex form of the objective function and the non-convex constraints. To deal with the max-min objective function, we first introduce an auxiliary variable
$\chi$, and the problem can be reformulated as
\begin{subequations}
\begin{align}
\max _{ \tilde{\boldsymbol{t}}} \quad &\chi \\
\text{s.t.}\quad  &\boldsymbol{a}(\theta)^{H} \boldsymbol{\Phi} \mathbf{H}(\tilde{\boldsymbol{t}})\mathbf{R} \mathbf{H}(\tilde{\boldsymbol{t}})^{H} \boldsymbol{\Phi}^{H} \boldsymbol{a}(\theta_l)\geq \chi , l\in \mathcal{L},\label{eq:x_sensing}\\
& \mathrm{SINR}_{k} \geq \Gamma,\forall k,\label{eq:x_SINR}\\ 
& \left\|\boldsymbol{t}_{k}-\boldsymbol{t}_{q}\right\|_{2} \geq D, k \neq q,\label{eq:x_distance}\\
& \tilde{\boldsymbol{t}} \in \mathcal{C}.\label{eq:transmit_region}
\end{align}\label{eq:x}%
\end{subequations}
The main challenges of solving Problem (\ref{eq:x}) lie in the complicated expressions of $\mathbf{H}(\tilde{\boldsymbol{t}})$, $\mathbf{h}_{1, k}(\tilde{\boldsymbol{t}})$ and the tight coupling of $\left\{\boldsymbol{t}_{m}\right\}_{m=1}^{M}$. To this end, we aim to solve Problem (\ref{eq:x}) in an alternating manner. Specifically, we   solve $M$ subproblems of (\ref{eq:x}), which respectively optimize one transmit  MA position $\boldsymbol{t}_{m}$, with all the
other variables being fixed. The developed alternating optimization algorithm can obtain a locally suboptimal
solution for Problem (\ref{eq:x}) by solving the above $M$
subproblems alternately.

We next consider the optimization of $\boldsymbol{t}_{m}$ with given $\left\{\boldsymbol{t}_{{q}}\right\}_{{q} \neq m} $. For constraint (\ref{eq:x_SINR}), we first rewrite the SINR constraint with respect to  the $k$-th user as\begin{equation}\label{eq:sinr_constraint_k}
\mathbf h_k^H\left[\left(1+\Gamma^{-1}\right) \mathbf R_k-\mathbf R\right] \mathbf h_k\geq \sigma^{2}_k.
\end{equation}

\noindent Since the antenna position $\boldsymbol t_m$ is only related to $\boldsymbol g(\boldsymbol{t}_m)$, we can convert constraint (\ref{eq:sinr_constraint_k}) to a more tractable form \begin{equation}
\left(\boldsymbol p_k^H \boldsymbol G(\tilde{\boldsymbol t})+ \boldsymbol q_k^H  \boldsymbol G_k(\tilde{\boldsymbol t})\right)\widetilde{\mathbf R}_k
\left(\boldsymbol G(\tilde{\boldsymbol t})^H \boldsymbol p_k+ \boldsymbol G_k(\tilde{\boldsymbol t})^H\boldsymbol q_k\right)\geq \sigma^{2}_k,\label{eq:t_SINR}
\end{equation}
where $\boldsymbol p_k^H=\mathbf{h}^H_{1, k}\boldsymbol{\Phi} \boldsymbol{F}(\boldsymbol{r})^{H} \boldsymbol{\Sigma}$ and $\boldsymbol q_k^H=\boldsymbol 1^H\boldsymbol{\Sigma}_k$ are invariant to $\tilde{\boldsymbol{t}}$. We then define $ x_{k,q}=\boldsymbol p_k^H \boldsymbol g(\boldsymbol{t}_q)+ \boldsymbol q_k^H   \boldsymbol g_k(\boldsymbol{t}_q),  q=1,2,\cdots,M$, constraint (\ref{eq:t_SINR}) can be recast as 
\begin{equation}\label{eq:sinr_constraint_2}
 \underbrace{\left[\widetilde{\mathbf R}_k \right]_{m,m}\left|x_{k,m}\right|^2+ 2\Re \{\widetilde{a}_k  x_{k,m}\boldsymbol  \}}_{\tilde{I}({\boldsymbol{t}_m})}\geq \sigma_k^2- \widetilde{b}_k,
\end{equation}
where $\tilde{a}_k$ and $\tilde{b}_k$ are given at the bottom of the next page. \begin{figure*}[b]
\hrulefill
\begin{equation}
 \widetilde{a}_k =\sum_{q=1,q\neq m}^{M}\left[\widetilde{\mathbf R}_k \right]_{m,q}  x_{k,q}^*
\end{equation}
\begin{equation}
\widetilde{b}_k = \sum_{p=1,p\neq m}^M \left( \sum_{q=1,q\neq m}^{p-1} 2\Re \{\left[\widetilde{\mathbf R}_k \right]_{p,q} x_{k,p} x_{k,q}^*\} + \left[\widetilde{\mathbf R}_k \right]_{p,p}\left|x_{k,p}\right|^2\right)
\end{equation}
\end{figure*}It is worth noting that $ x_{k,m}$ is the optimization variable related to the antenna position $\boldsymbol t_m$, while  $\widetilde{a}_k$ and $\widetilde{b}_k$ are variables invariant to   $\boldsymbol t_m$. To fully reveal the optimization variable $\boldsymbol{t}_m$ in the current form of constraint (\ref{eq:sinr_constraint_2}), we expand the left-hand-side of constraint (\ref{eq:sinr_constraint_2}) in (\ref{eq:expansion_of_It}), as shown at the bottom of the next page. In (\ref{eq:expansion_of_It}), $\boldsymbol{P}_k\triangleq\boldsymbol{p}_k\boldsymbol{p}_k^H$, $\boldsymbol{Q}_k\triangleq\boldsymbol{q}_k\boldsymbol{q}_k^H$. 

\begin{figure*}[b]
\hrulefill
\begin{equation}\label{eq:expansion_of_It}
\begin{aligned}
\tilde{I}({\boldsymbol{t}_m})
&=\left[\widetilde{\mathbf R}_k \right]_{m,m}\left(\boldsymbol{g}_k(\boldsymbol{t}_m)^H\boldsymbol{Q}_k\boldsymbol{g}_k(\boldsymbol{t}_m)+\boldsymbol{g}(\boldsymbol{t}_m)^H\boldsymbol{P}_k\boldsymbol{g}(\boldsymbol{t}_m)+2\Re\{\boldsymbol{p}_k^H\boldsymbol{g}(\boldsymbol{t}_m)\boldsymbol{g}_k(\boldsymbol{t}_m)^H\boldsymbol{q}_k\}\right)+2\Re\{\tilde{a}_k\boldsymbol{p}_k^H\boldsymbol{g}(\boldsymbol{t}_m)+\tilde{a}_k\boldsymbol{q}_k^H\boldsymbol{g}_k(\boldsymbol{t}_m)\}\\
&=2\sum_{i=1}^{L_t^k-1}\sum_{j=i+1}^{L_t^k}2\left[\widetilde{\mathbf R}_k \right]_{m,m}\left|\left[\boldsymbol{Q}_k\right]_{i,j}\right|\cos\left(\frac{2\pi}{\lambda}\left(\rho_{t,k}^i(\boldsymbol{t}_m)-\rho_{t,k}^j(\boldsymbol{t}_m)\right)+\angle\left[\boldsymbol{Q}_k\right]_{i,j}\right)+\sum_{i=1}^{L_t^k}\left[\widetilde{\mathbf R}_k \right]_{m,m}\left[\boldsymbol{Q}_k\right]_{i,i}\\
&\quad+2\sum_{i=1}^{L_t-1}\sum_{j=i+1}^{L_t}\left[\widetilde{\mathbf R}_k \right]_{m,m}\left|\left[\boldsymbol{P}_k\right]_{i,j}\right|\cos\left(\frac{2\pi}{\lambda}\left(\rho_{t}^i(\boldsymbol{t}_m)-\rho_{t}^j(\boldsymbol{t}_m)\right)+\angle\left[\boldsymbol{P}_k\right]_{i,j}\right)+\sum_{i=1}^{L_t}\left[\widetilde{\mathbf R}_k \right]_{m,m}\left[\boldsymbol{P}_k\right]_{i,i}\\
&\quad+2\sum_{i=1}^{L_t}\sum_{j=1}^{L_t^k}\left[\widetilde{\mathbf R}_k \right]_{m,m}\left|\left[
\boldsymbol{p}_k\right]_i\right|\left|\left[
\boldsymbol{q}_k\right]_j\right|\cos\left(\frac{2\pi}{\lambda}\left(\rho_{t}^i(\boldsymbol{t}_m)-\rho_{t,k}^j(\boldsymbol{t}_m)\right)+\angle \left[\boldsymbol{q}_k\right]_j-\angle \left[\boldsymbol{p}_k\right]_i\right)\\
&\quad+2\sum_{i=1}^{L_t}\left|
\tilde{a}_k\right|\left|\left[
\boldsymbol{p}_k\right]_i\right|\cos\left(\frac{2\pi}\lambda\rho_{t}^i(\boldsymbol{t}_m)-\angle\left[
\boldsymbol{p}_k\right]_i+\angle\tilde{a}_k\right)+2\sum_{j=1}^{L_t^k}\left|
\tilde{a}_k\right|\left|\left[
\boldsymbol{q}_k\right]_j\right|\cos\left(\frac{2\pi}\lambda\rho_{t,k}^i(\boldsymbol{t}_m)-\angle\left[
\boldsymbol{q}_k\right]_j+\angle\tilde{a}_k\right)
\end{aligned}
\end{equation}

\end{figure*}

However, $\tilde{I}(\boldsymbol{t}_m)$ is neither concave nor convex with respect to
$\boldsymbol{t}_m$,
making constraint (\ref{eq:sinr_constraint_2}) still non-convex and thus intractable.  To tackle this problem, we use the SCA method and transform the non-convex constraint. Specifically, by using the Taylor’s theorem, we can construct a quadratic surrogate function which serves as a strict convex constraint to (\ref{eq:sinr_constraint_2}). With given local point $\boldsymbol{t}_m^{(i)}\in \mathbb{R}^2$ in the $i$-th iteration, a lower bound can be obtained as $\tilde{I}\left(\boldsymbol{t}_m\right) \geq\tilde{I}\left(\boldsymbol{t}_m^{(i)}\right)+\nabla \tilde{I}\left(\boldsymbol{t}_m^{(i)}\right)^{T}\left(\boldsymbol{t}_m-\boldsymbol{t}_m^{(i)}\right)-\frac{\tilde{\delta}_{m}^{(i)}}{2}\left(\boldsymbol{t}_m-\boldsymbol{t}_m^{(i)}\right)^{T}\left(\boldsymbol{t}_m-\boldsymbol{t}_m^{(i)}\right)$, where  $\tilde{\delta}_{m}^{(i)}$ is a positive real number which satisfies $\tilde{\delta}_{m}^{(i)}\mathbf{I}_2\succeq\nabla^2\tilde{I}(\boldsymbol{t}_m)$.  The gradient vector $\nabla\tilde{I}(\boldsymbol{t}_m)$ and the  Hessian matrix $\nabla^{2}\tilde{I}\left(\boldsymbol{t}_{m}\right)$ are given in Appendix A. According to Appendix A and $\left\|\nabla^{2} \tilde{I}\left(\boldsymbol{t}_{m}\right)\right\|_{2} \mathbf{I}_{2} \succeq \nabla^{2} \tilde{I}\left(\boldsymbol{t}_m\right)$, we can construct $\tilde{\delta}_m^{(i)}$ as follows:
\begin{equation}
\begin{aligned}
\tilde{\delta}_m^{(i)}&=\frac{64\pi^2}\lambda \left[\sum_{i=1}^{L_t^k-1}\sum_{j=i+1}^{L_t^k}\left[\widetilde{\mathbf R}_k \right]_{m,m}\left|\left[\boldsymbol{Q}_k\right]_{i,j}\right|\right.\\
&\quad +\sum_{i=1}^{L_t-1}\sum_{j=i+1}^{L_t}\left[\widetilde{\mathbf R}_k \right]_{m,m}\left|\left[\boldsymbol{P}_k\right]_{i,j}\right|\\
&\quad+\left.\sum_{i=1}^{L_t}\sum_{j=1}^{L_t^k}\left[\widetilde{\mathbf R}_k \right]_{m,m}\left|\left[\boldsymbol{p}_k\right]_i\right|\left|\left[ \boldsymbol{q}_k\right]_j\right|\right]\\
&\quad+\frac{16\pi^2}\lambda\left[\sum_{i=1}^{L_t}\left|\tilde{a}_k\right|\left|\left[ \boldsymbol{p}_k\right]_i\right|+\sum_{j=1}^{L_t^k}\left|\tilde{a}_k\right|\left|\left[\boldsymbol{q}_k\right]_j\right|\right]\\
&\geq \left\|\nabla^{2} \bar{I}\left(\boldsymbol{t}_{m}\right)\right\|_{F} \overset{(a)}{\geq} \left\|\nabla^{2} \bar{I}\left(\boldsymbol{t}_{m}\right)\right\|_{2},
\end{aligned}
\end{equation}
which satisfies  $\tilde{\delta}_m^{(i)} \boldsymbol{I}_{2} \succeq \left\|\nabla^{2} \tilde{I}\left(\boldsymbol{t}_{m}\right)\right\|_{2}  \boldsymbol{I}_{2}\succeq\nabla^{2} \tilde{I}\left(\boldsymbol{t}_{m}\right)$. The inequality marked by $(a)$ holds because $\|\nabla^{2} \tilde{I}\left(\boldsymbol{t}_{m}\right)\|_{F}=\sqrt{\sum_{i=1} \sigma_{i}^{2}(\nabla^{2} \tilde{I}\left(\boldsymbol{t}_{m}\right))}\geq \sigma_{\max }(\nabla^{2} \tilde{I}\left(\boldsymbol{t}_{m}\right))=\|\nabla^{2} \tilde{I}\left(\boldsymbol{t}_{m}\right)\|_2$, where 
$\sigma _{i}(\nabla^{2}\tilde{I}\left(\boldsymbol{t}_{m}\right))$ are the singular values of $\nabla^{2} \tilde{I}\left(\boldsymbol{t}_{m}\right)$.
Thus we can rewrite constraint (\ref{eq:sinr_constraint_2}) as 
\begin{align}\label{eq:SINR_SCA}
\nonumber\tilde{I}\left(\boldsymbol{t}_m\right) &\geq\tilde{I}\left(\boldsymbol{t}_m^{(i)}\right)+\nabla \tilde{I}\left(\boldsymbol{t}_m^{(i)}\right)^{T}\left(\boldsymbol{t}_m-\boldsymbol{t}_m^{(i)}\right)\\
\nonumber&\quad -\frac{\tilde{\delta}_m^{(i)}}{2}\left(\boldsymbol{t}_m-\boldsymbol{t}_m^{(i)}\right)^{T}\left(\boldsymbol{t}_m-\boldsymbol{t}_m^{(i)}\right)\nonumber\\
\nonumber&=-\frac{\tilde{\delta}_m^{(i)}}{2} \boldsymbol{t}_m^{T} \boldsymbol{t}_m+\left(\nabla \tilde{I}\left(\boldsymbol{t}_m^{(i)}\right)+\tilde{\delta}_m^{(i)} \boldsymbol{t}_m^{(i)}\right)^{T} \boldsymbol{t}_m\\
\nonumber&\quad+\underbrace{\tilde{I}\left(\boldsymbol{t}_m^{(i)}\right)-\frac{\tilde{\delta}_m^{(i)}}{2}\left(\boldsymbol{t}_m^{(i)}\right)^{T} \boldsymbol{t}_m^{(i)}-\nabla \tilde{I}\left(\boldsymbol{t}_m^{(i)}\right)^{T}\boldsymbol{t}_m^{(i)}}_{\mathrm{const}}\\\
&\geq \sigma^2- \widetilde{b}_k.
\end{align}
Next, we tackle the non-convexity of constraint (\ref{eq:x_sensing}). Define  $\boldsymbol{d}(\theta_l)^{H}=\boldsymbol{a}(\theta_l)^{H} \boldsymbol{\Phi}\boldsymbol{F}(\boldsymbol{r})^{H} \boldsymbol{\Sigma}$, and constraint  (\ref{eq:x_sensing}) can be rewritten as \begin{align}\label{eq:Sensing_sca}
& \quad \boldsymbol{a}(\theta_l)^{H} \boldsymbol{\Phi} \mathbf{H}(\tilde{\boldsymbol{t}})\mathbf{R} \mathbf{H}(\tilde{\boldsymbol{t}})^{H} \boldsymbol{\Phi}^{H} \boldsymbol{a}(\theta_l) \nonumber\\
&=\underbrace{\boldsymbol{d}(\theta_l)^{H} \boldsymbol{A}(\boldsymbol{t}_m)\boldsymbol{d}(\theta_l)+\left[\mathbf R\right]_{m,m}|\boldsymbol{d}(\theta_l)^{H}\boldsymbol{g}(\boldsymbol{t}_m)|^2}_{\bar{I}(\boldsymbol{t}_m)}\nonumber\\
&\quad +\boldsymbol{d}(\theta_l)^{H}\boldsymbol{B}_m  \boldsymbol{d}(\theta_l)\geq \chi,
\end{align}where $ \boldsymbol{A}(\boldsymbol{t}_m)$ is a linear function of $\boldsymbol{g}(\boldsymbol{t}_m)$ and $\boldsymbol{B}_m$ is a constant matrix independent of $\boldsymbol{t}_m$, which are defined in (\ref{A(t)}) and (\ref{B})  respectively at the bottom of the next page. Since $I(\boldsymbol{t}_m)$ can be lower bounded by its first-order
Taylor expansion, we apply the SCA method and the non-convex constraint (\ref{eq:Sensing_sca}) can be
rewritten as
\begin{figure*}[b]
\hrulefill
\begin{equation}\label{A(t)}
 \boldsymbol{A}(\boldsymbol{t}_m) =\sum_{q=1,q\neq m}^{M}\left(\left[\mathbf{R}\right]_{m,q} \boldsymbol{g}(\boldsymbol{t}_m)\boldsymbol{g}(\boldsymbol{t}_q)^H+\left[\mathbf{R}\right]_{m,q}^{*} \boldsymbol{g}(\boldsymbol{t}_q)\boldsymbol{g}(\boldsymbol{t}_m)^H\right) 
\end{equation}

\begin{equation}\label{B}
\boldsymbol{B}_m=    \sum_{k=1,k\neq m}^M \left( \sum_{q=1,q\neq m}^{k-1} \left( \left[\mathbf{R}\right]_{k,q} \boldsymbol{g}(\boldsymbol{t}_k)\boldsymbol{g}(\boldsymbol{t}_q)^H +\left[\mathbf{R}\right]_{k,q}^{* } \boldsymbol{g}(\boldsymbol{t}_q)\boldsymbol{g}(\boldsymbol{t}_k)^H \right) +\left[\mathbf{R}\right]_{k,k}\boldsymbol{g}(\boldsymbol{t}_k)\boldsymbol{g}(\boldsymbol{t}_k)^H\right)  
\end{equation}
\end{figure*}

\begin{align}\label{eq:Sensing_first_taylor}
&\underbrace{2 \left[\mathbf R\right]_{m,m}\Re\left\{\boldsymbol{g}(\boldsymbol{t}_m^{(i)})^H \boldsymbol{d}(\theta_{l}) \boldsymbol{d}(\theta_{l})^{H} \boldsymbol{g}(\boldsymbol{t}_m)\right\}+\boldsymbol{d}(\theta_l)^{H} \boldsymbol{A}(\boldsymbol{t}_m)\boldsymbol{d}(\theta_l)}_{\bar{I}(\boldsymbol{t}_m)}\nonumber\\
&\geq \chi-\boldsymbol{d}(\theta_l)^{H}\boldsymbol{B}_m  \boldsymbol{d}(\theta_l)+\left[\mathbf R\right]_{m,m}\left|\boldsymbol{d}(\theta_{l})^{H} \boldsymbol{g}(\boldsymbol{t}_m^{(i)})\right|^{2}.
\end{align}
Here $\bar{I}(\boldsymbol{t}_m)$ is a linear function of $\boldsymbol{g}(\boldsymbol{t}_m)$, though it is still neither concave nor convex over $\boldsymbol{t}_m$.  Analogous  to (\ref{eq:SINR_SCA}), we can construct a second-order Taylor expansion-based concave
lower bound for $\bar{I}(\boldsymbol{t}_m)$. For ease of exposition, we define $\boldsymbol{b}_1^T \triangleq 2 \left[\mathbf R\right]_{m,m}\boldsymbol{g}(\boldsymbol{t}_m^{(i)})^H \boldsymbol{d}_{l}(\theta) \boldsymbol{d}_{l}(\theta)^{H}$, $\boldsymbol{b}_2^T \triangleq 2\sum_{q=1,q\neq m}^{M}\left(\left[\mathbf{R}\right]_{m,q} \boldsymbol{g}(\boldsymbol{t}_q)^H\right) \boldsymbol{d}(\theta_l)\boldsymbol{d}(\theta_l)^{H}$ and $\boldsymbol{b}=\boldsymbol{b}_1+\boldsymbol{b}_2$. Thus $\bar{I}(\boldsymbol{t}_m)$ can be  rewritten as 
\begin{align}
\nonumber\bar{I}(\boldsymbol{t}_m) & =\Re\left\{2 \left[\mathbf R\right]_{m,m}\boldsymbol{g}(\boldsymbol{t}_m^{(i)})^H \boldsymbol{d}(\theta_{l}) \boldsymbol{d}(\theta_{l})^{H} \boldsymbol{g}(\boldsymbol{t}_m)\right\}\\\nonumber
&\quad+\Re\left\{2\boldsymbol{d}(\theta_l)^{H} \left(\sum_{q=1,q\neq n}^{M}\left[\mathbf R\right]_{m,q} \boldsymbol{g}(\boldsymbol{t}_q)^H\right) \boldsymbol{d}(\theta_l)\boldsymbol{g}(\boldsymbol{t}_m)\right\}\\ \nonumber
&=\Re\left\{\boldsymbol{b}_1^T\boldsymbol{g}(\boldsymbol{t}_m)+\boldsymbol{b}_2^T\boldsymbol{g}(\boldsymbol{t}_m)\right\}\\ \nonumber
&=\Re\left\{\boldsymbol{b}^T\boldsymbol{g}(\boldsymbol{t}_m)\right\}\\
&=\sum_{k=1}^{L_{t}}\left|\left[\boldsymbol b\right]_k\right| \cos \left(\varrho  ^{k}\left(\boldsymbol{t}_{m}\right)\right),
\end{align}
where $\varrho^{k}\left(\boldsymbol{t}_{m}\right) \triangleq2 \pi \rho_{t}^{k}\left(\boldsymbol{t}_{m}\right) / \lambda+\angle \left[\boldsymbol b\right]_k$.  In a similar way, we can calculate the the gradient vector $\nabla \bar{I}\left(\boldsymbol{t}_m\right)$ and  the Hessian matrix $\nabla^2\bar{I}(\boldsymbol{t}_m)$ of $\bar{I}(\boldsymbol{t}_m)$, which are omitted for simplicity. Hence, we can obtain the $\bar{\delta}_m^{(i)}$ as follows:
\begin{equation}\label{eq:Sensing_SCA}
\bar{\delta}_m^{(i)}  = \frac{8 \pi^{2}}{\lambda^{2}} \sum_{k  = 1}^{L_{r}}\left|\left[\boldsymbol{b}\right]_{k}\right| ,
\end{equation}
which satisfies  $\bar{\delta}_m^{(i)} \boldsymbol{I}_{2} \succeq \left\|\nabla^{2} \bar{I}\left(\boldsymbol{t}_{m}\right)\right\|_{2}  \boldsymbol{I}_{2}\succeq\nabla^{2} \bar{I}\left(\boldsymbol{t}_{m}\right)$. Thus, we can rewrite constraint (\ref{eq:Sensing_first_taylor}) as 

\begin{align}
\nonumber\bar{I}\left(\boldsymbol{t}_m\right) &\geq\bar{I}\left(\boldsymbol{t}_m^{(i)}\right)+\nabla \bar{I}\left(\boldsymbol{t}_m^{(i)}\right)^{T}\left(\boldsymbol{t}_m-\boldsymbol{t}_m^{(i)}\right)\\\nonumber
&\quad -\frac{\delta_{n}^{(i)}}{2}\left(\boldsymbol{t}_m-\boldsymbol{t}_m^{(i)}\right)^{T}\left(\boldsymbol{t}_m-\boldsymbol{t}_m^{(i)}\right)\\\nonumber
&=-\frac{\delta_{n}^{(i)}}{2} \boldsymbol{t}_m^{T} \boldsymbol{t}_m+\left(\nabla \bar{I}\left(\boldsymbol{t}_m^{(i)}\right)+\delta_{n}^{(i)} \boldsymbol{t}_m^{(i)}\right)^{T} \boldsymbol{t}_m\\\nonumber
&\quad+\underbrace{\bar{I}\left(\boldsymbol{t}_m^{(i)}\right)-\frac{\delta_{n}^{(i)}}{2}\left(\boldsymbol{t}_m^{(i)}\right)^{T} \boldsymbol{t}_m^{(i)}-\nabla \bar{I}\left(\boldsymbol{t}_m^{(i)}\right)^{T}\boldsymbol{t}_m^{(i)}}_{\mathrm{const}}\\
&\geq \chi-\boldsymbol{d}(\theta_l)^{H}\boldsymbol{B}_m  \boldsymbol{d}(\theta_l)+\left[\mathbf R\right]_{m,m}\left|\boldsymbol{d}_{l}(\theta)^{H} \boldsymbol{g}(\boldsymbol{t}_m^{(i)})\right|^{2}.\label{eq:sensing_taylor}
\end{align}
 Finally, constraint (\ref{eq:x_distance}) can be relaxed at the given point $\boldsymbol{t}_m^{(i)}$  as follows according to \cite{ma2023mimo}
\begin{equation}
\left\|\boldsymbol{t}_{m}-\boldsymbol{t}_{q}\right\|_{2} \geq \frac{1}{\left\|\boldsymbol{t}_{m}^{(i)}-\boldsymbol{t}_{q}\right\|_{2}}\left(\boldsymbol{t}_{m}^{(i)}-\boldsymbol{t}_{q}\right)^{T}\left(\boldsymbol{t}_{m}-\boldsymbol{t}_{q}\right)\geq D.\label{eq:mim_dis}
\end{equation}
Hence, the antenna position design problem can be reformulated as
\begin{subequations}
\begin{align}
\max _{ \boldsymbol{t}_m}  \quad & \chi  \\
\text{s.t.} \quad  & \text{(\ref{eq:transmit_region}), (\ref{eq:SINR_SCA}), (\ref{eq:sensing_taylor}), (\ref{eq:mim_dis})},
\end{align}\label{eq:t_final}%
\end{subequations}which is a convex quadratically constrained problem (QCP) that can be solved efficiently by standard convex solvers such as CVX \cite{boyd2004convex}.

\begin{algorithm}[!t]
	\caption{Alternating Optimization for Solving Problem (\ref{eq:op_problem})}
	\label{alg2}
	\begin{algorithmic}[1]
		\STATE \emph{Input:} $\boldsymbol{\Sigma}$, $P_0$, $\sigma$, $M$, $N$, $L_r$, $L_t$, $\{\theta^r_i\}_{i=1}^{L_r}$, $\{\phi^r_i\}_{i=1}^{L_r}$, $\{\theta^t_j\}_{j=1}^{L_t}$, $\{\phi^t_j\}_{j=1}^{L_t}$,  $\mathcal{C}$, $D$, $\epsilon_1$, $\epsilon_2$.
        \STATE Initialize $\{\boldsymbol{t}_m\}_{n=1}^M$.
        \WHILE{Increase of the minimum beampattern gain in (\ref{eq:beampattern}) is above $\epsilon$}
            \STATE Obtain the optimal solution of $\mathbf{R}$  with given  $\tilde{\boldsymbol{t}}$ and $\mathbf \Phi$ by solving Problem (\ref{eq:BS_beamforming}).
            \STATE Obtain the optimal solution of $\mathbf\Phi$  with given  $\tilde{\boldsymbol{t}}$ and $\mathbf R$ by solving Problem (\ref{eq:RIS}).
            \FOR {$n=1\rightarrow M$}
                \WHILE{Increase of the minimum beampattern gain in (\ref{eq:beampattern}) is above $\epsilon$}               
                    \STATE Given $\mathbf{R}$, $\mathbf{\Phi}$, solve Problem (\ref{eq:t_final}) to update $\boldsymbol{t}_m$.
                \ENDWHILE
            \ENDFOR          
        \ENDWHILE

        \STATE \emph{Output:} $\mathbf{R}$, $\mathbf{\Phi}$, $\tilde{\boldsymbol{t}}$.
	\end{algorithmic}
\end{algorithm}

\subsection{Convergence and Complexity Analysis}
Based on the above discussions, the detailed procedure of the overall AO-based algorithm are summarized in Algorithm 2. It is readily verified that the optimal value  of Problem (\ref{eq:op_problem}) monotonically increases in each step of Algorithm 2. Furthermore,  the optimal value is upper-bounded. As a result, Algorithm 2 is assured to converge.

Next, we analyze the computational complexity of Algorithm 1. In each iteration, the complexity mainly stems from solving the three subproblems. Firstly, given a solution accuracy $\epsilon$, the worst-case complexity to solve the SDP Problem (\ref{eq:BS_beamforming_SDR}) is $\mathcal{O}\left(M^{4.5}\log(1/\epsilon)\right)$\cite{Xia2020joint}. Let $T_1^\mathrm{max}$ and $T_2^\mathrm{max}$ denote the  maximum number of  inner-layer iterations  to update $\bf \Phi$ and $\tilde{\boldsymbol{t}}$, respectively. Then, the complexity of updating $\bf \Phi$ is  given by $\mathcal{O}\left(T_1^\mathrm{max}M^4\sqrt{N}\log(1/\epsilon)\right)$\cite{zuo2023exploi}, and the complexity of updating $\tilde{\boldsymbol{t}}$ is given by $\mathcal{O}\left(T_2^\mathrm{max}M^{3.5}\log(1/\epsilon)\right)$\cite{yu2023active}. Finally, the total complexity of the overall AO-based algorithm is given by $\mathcal{O}\left(T_3^{\mathrm{max}}\left(M^{4.5}+T_1^\mathrm{max}M^4\sqrt{N}+T_2^\mathrm{max}M^{3.5}\right)\log(1/\epsilon)\right)$, where $T_3^{\mathrm{max}}$  is the maximum number of outer-layer iterations of the overall algorithm.

\section{SIMULATION RESULTS}

In this section, we present simulation results to assess
the performance of the proposed  MA and RIS-aided ISAC
systems. In the simulation setup, we assume that the BS is fixed at $(0, 0)$ and RIS is located at $(12\,\mathrm m, 16\,\mathrm m)$. The users are randomly distributed in the rectangle area between $(20\, \mathrm m,0)$ and $(40\,\mathrm m,-20 \,\mathrm m)$. We consider the geometry channel model, where the numbers of transmit and receive paths are the same, i.e., $L_t = L_r=L=4$. According to the aforementioned system model, we consider LoS channel between the BS-RIS link and RIS-User link. In addition, the PRMs of the LoS channels are modeled as $\boldsymbol{\Sigma}[1,1]\sim\mathcal{CN}(0,K_0(\frac{d}{d_0})^{-\alpha}\kappa/(\kappa+1))$ and $\boldsymbol{\Sigma}[p,p]\sim\mathcal{CN}(0,K_0(\frac{d}{d_0})^{-\alpha}/((\kappa+1)(L-1))),\,p=2,3,\ldots,L$, where $\kappa$ denotes the ratio of the average power for LoS paths to that for NLoS paths. The distance-dependent path-loss is modeled as $K_0(\frac{d}{d_0})^{-\alpha}$, where $K_{0}=-40$ dB is the average channel power gain at the reference distance $d_0=1\,\mathrm m$ and $\alpha$ is the pathloss exponent.  Since  obstacles exist between the BS-User links, we consider Rayleigh fading and an additional shadow fading with a standard deviation of $15$ dB between the NLoS channels.  Unless otherwise stated, the simulation parameters are set as follows: The number of  MAs at the DFRC BS  $M = 8$, the number of downlink users $K = 2$, the interested angles   $\{\theta_{1},\theta_{2},\cdots,\theta_{5}\}=\{-60^{\circ},-30^{\circ},0^{\circ},30^{\circ},60^{\circ}\}$, the transmit power at the DFRC BS  $P_{0}=40$ dBm, the user SINR threshold $\Gamma=10$ dB, the noise power at the receive antenna $\sigma^2_k=-80$ dBm, the number  paths $L=4$, and the pathloss exponents of BS-RIS, RIS-User, BS-User channels  $\alpha=2.5, 2.5, 3.5$, respectively. All the results are averaged over 400 independent channel realizations.

\subsection{Convergence Behavior of the Proposed Algorithms}

\begin{figure}
        \centering
        \includegraphics[width=1.05\linewidth]{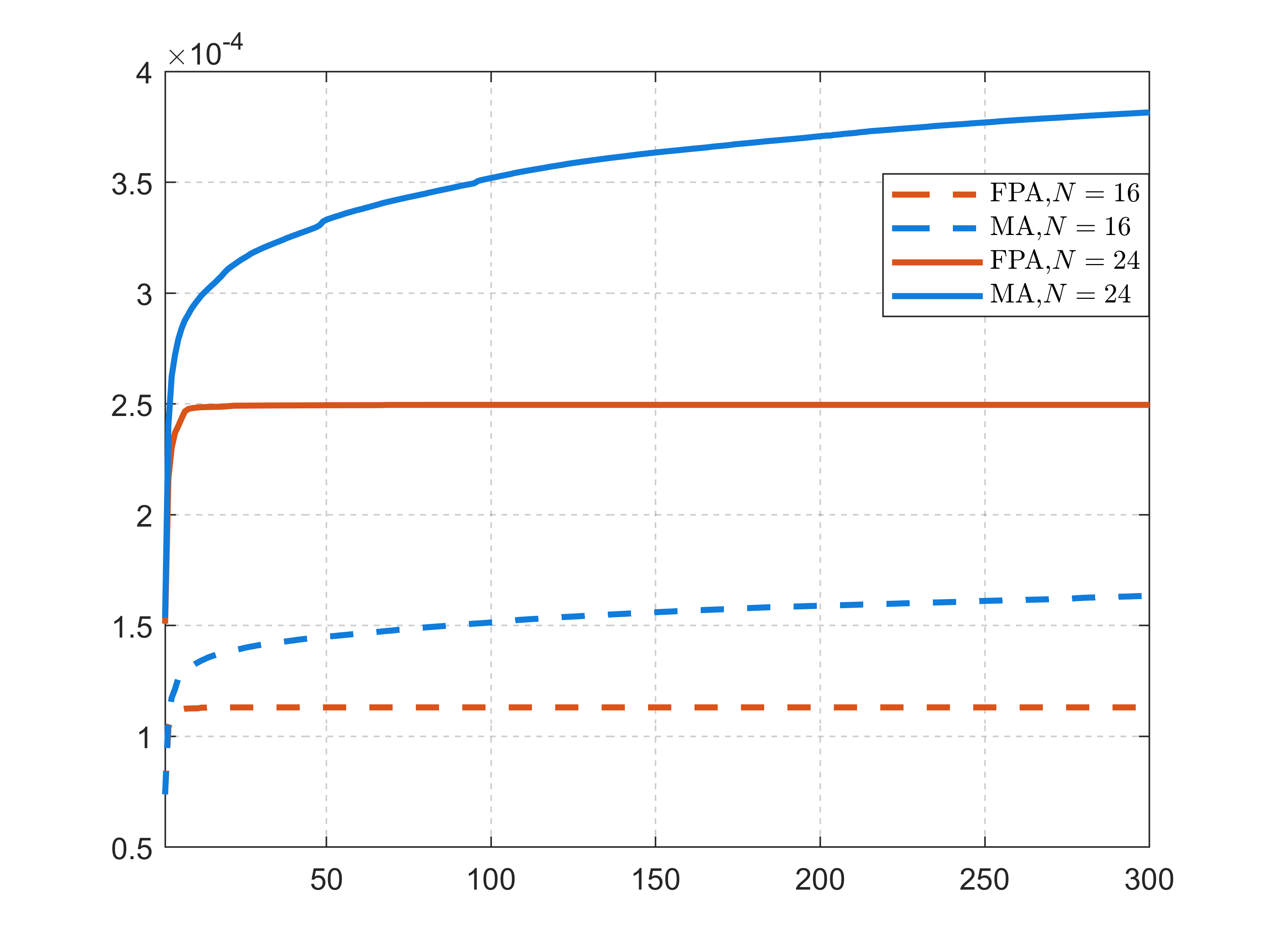}
        \caption{Convergence behavior.}
        \label{fig:convergence}
\end{figure}

The convergence behavior of the proposed algorithm is shown in Fig. \ref{fig:convergence}. As  illustrated, the minimum beampattern gains of all schemes increase with the iteration index and converge within 300 iterations, thus validating the effectiveness of our proposed algorithm. In addition, the  MA-aided schemes outperform their FPA-aided counterparts, but require more iterations to converge. This is owing to the procedure of the antenna position optimization in our proposed algorithm.

\subsection{Beampattern at the RIS }

\begin{figure}
    \centering
    \includegraphics[width=1.05\linewidth]{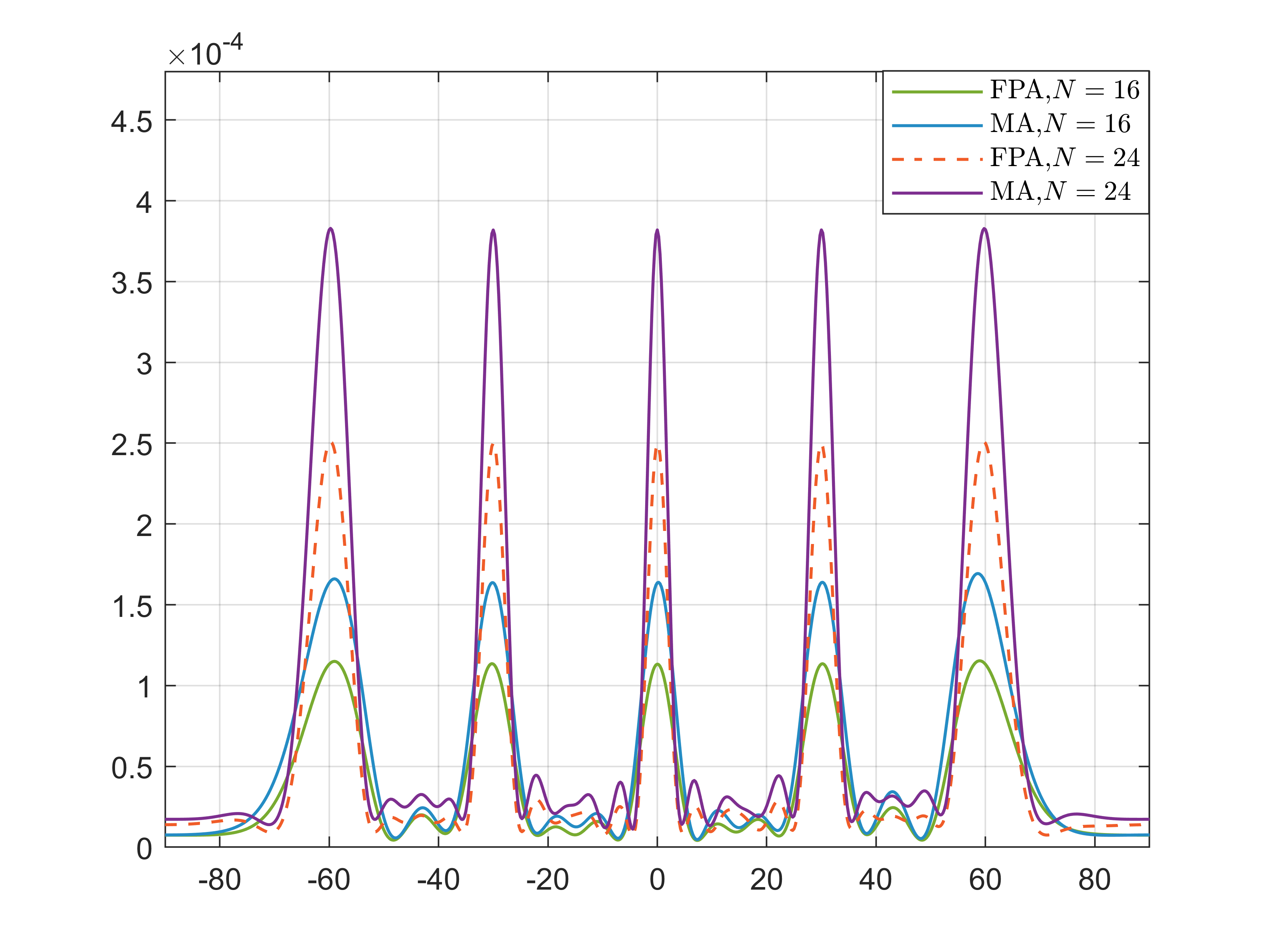}
    \caption{The beampattern gains of different schemes.}
    \label{fig:beampattern}
\end{figure}

Fig. \ref{fig:beampattern} illustrates the beampattern gain of the different schemes with $\theta\in \left[-90^\circ,90^\circ\right]$. As depicted in Fig. \ref{fig:beampattern}, the values of beampattern gain in the direction of the interested angles in each scheme are approximately the same, which can be attributed to the principle of fairness inherent in the max-min optimization problem. In addition, the beampattern gain of the  MA-aided schemes exceed their FPA-aided counterpart by approximately 30\%, which reveals the advantage of the deployment of  MAs at the DFRC BS, owing to its capacity of channel state configuration and  array geometry reconstruction. Furthermore, the increase in the number of the RIS reflecting elements will result in narrower main lobe of the beampattern, which means better directivity of the sensing function.

\subsection{Channel Characteristics under Antenna Position Optimization}

\begin{figure}
    \centering

    \includegraphics[width=1.05\linewidth]{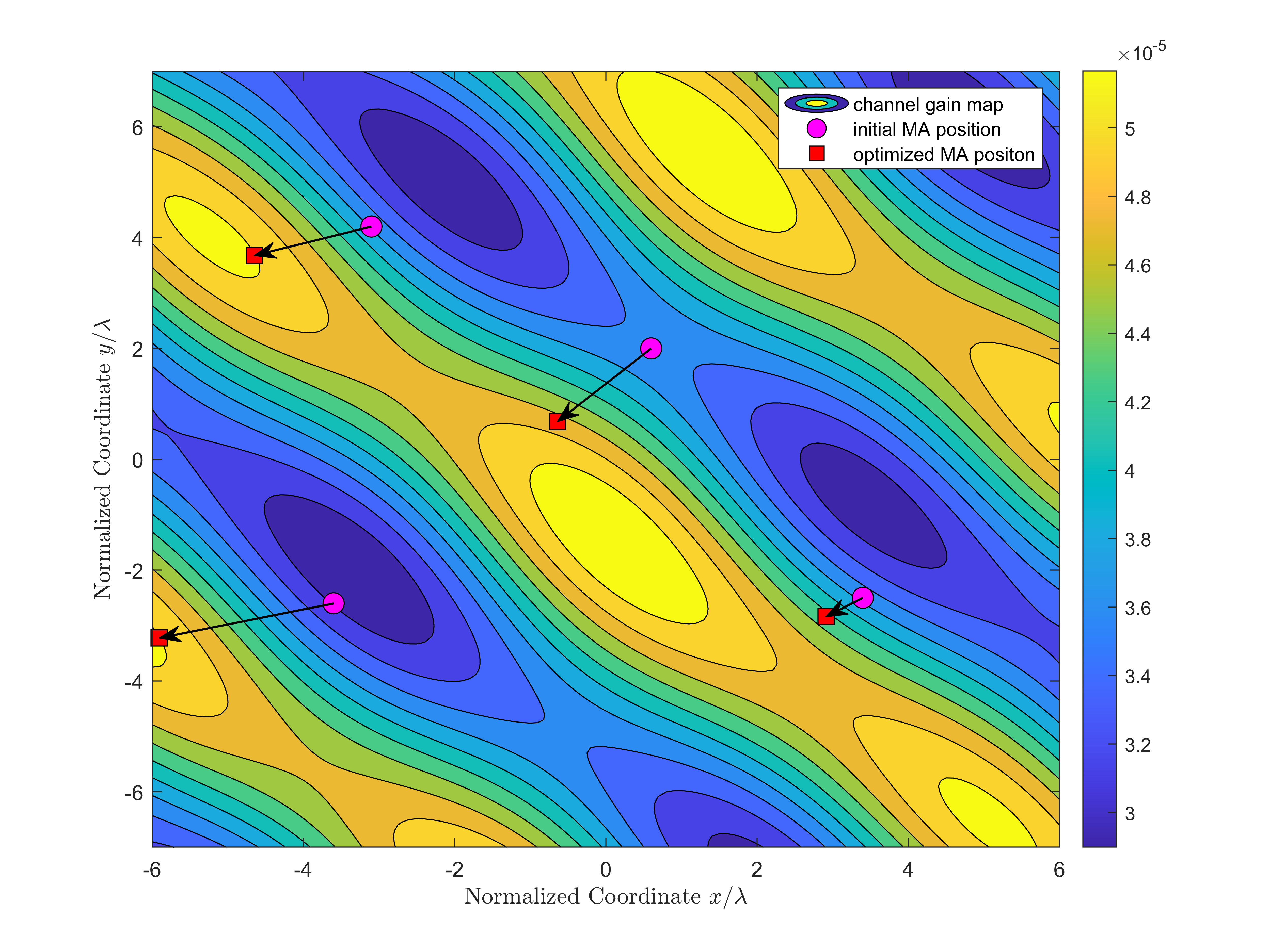}        
    \caption{Example for the channel power gain in the transmit region $\mathcal{C}$: Movement of $M=4$ antennas.}
    \label{fig:channel_example}
\end{figure}

\begin{figure}
    \centering
    \includegraphics[width=1.02\linewidth]{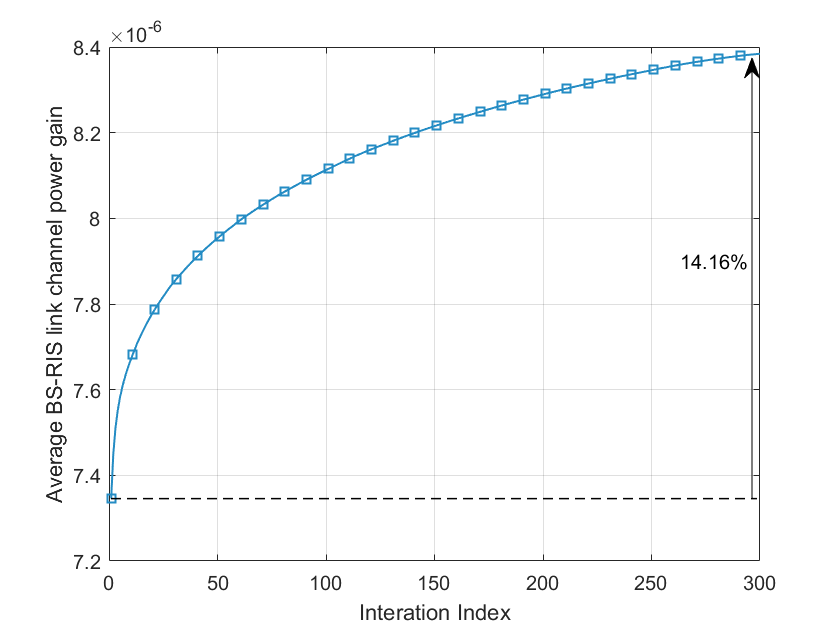}
    \caption{Channel gain of BS-RIS link  versus iteration index}
    \label{fig:BS-RIS}
\end{figure}

To explore the effect of  MA  on improving the  channel conditions and array geometry, one implementation in which the channel gain of BS-RIS link versus the  MAs' positions are presented in Fig. \ref{fig:channel_example}.  For convenience and clarity, we set the number of antennas as $M=4$, and the RIS sensing angles $\{\theta_1, \cdots, \theta_L\}$ are set to $\{-30^\circ, 0^\circ, 30^\circ\}$ with a service user number $K = 2$, and the channel power  gain of the BS-RIS link is defined as $\|\mathbf{H}(\tilde{\boldsymbol{t}})\|^2$. In the implementation, the channel gain $\|\mathbf{H}(\tilde{\boldsymbol{t}})\|^2$ is increased by 12\%, which is significantly smaller than the average system performance gain as depicted in Fig. \ref{fig:beampattern}. Furthermore, it can be observed that the positions of the  MAs are not all optimized to the locations with the maximum channel gain. This is reasonable because the ISAC system needs to simultaneously perform sensing and communication, and the antenna arrays at certain positions can more easily form the desired waveform pattern\cite{ma2024movable}. In addition, Fig. \ref{fig:BS-RIS} shows the variation of the average channel  gain  of the BS-RIS link during the process of antenna position optimization.  It can be observed that the channel power gain gradually improves with the number of iterations, once again verifying the role of  MA in improving the communication environment by mitigating small-scale fading caused by multipath effects.

\begin{figure}
    \centering
    \includegraphics[width=1.01\linewidth]{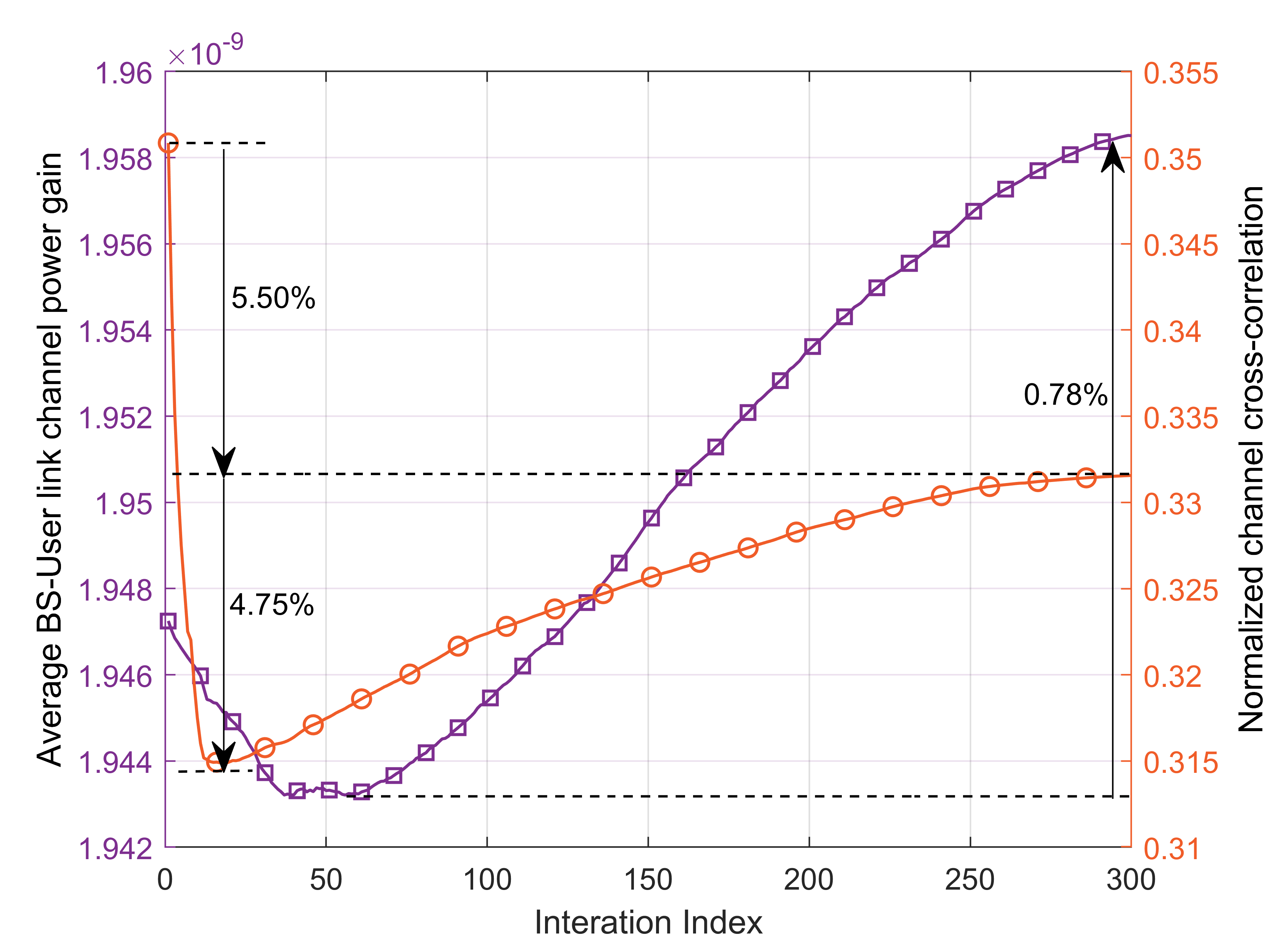}
    \caption{Average user channel power gain and normalized cross-correlation versus iteration index.}
    \label{fig:User_channel}
\end{figure}

\begin{figure}
    \centering
    \includegraphics[width=1.02\linewidth]{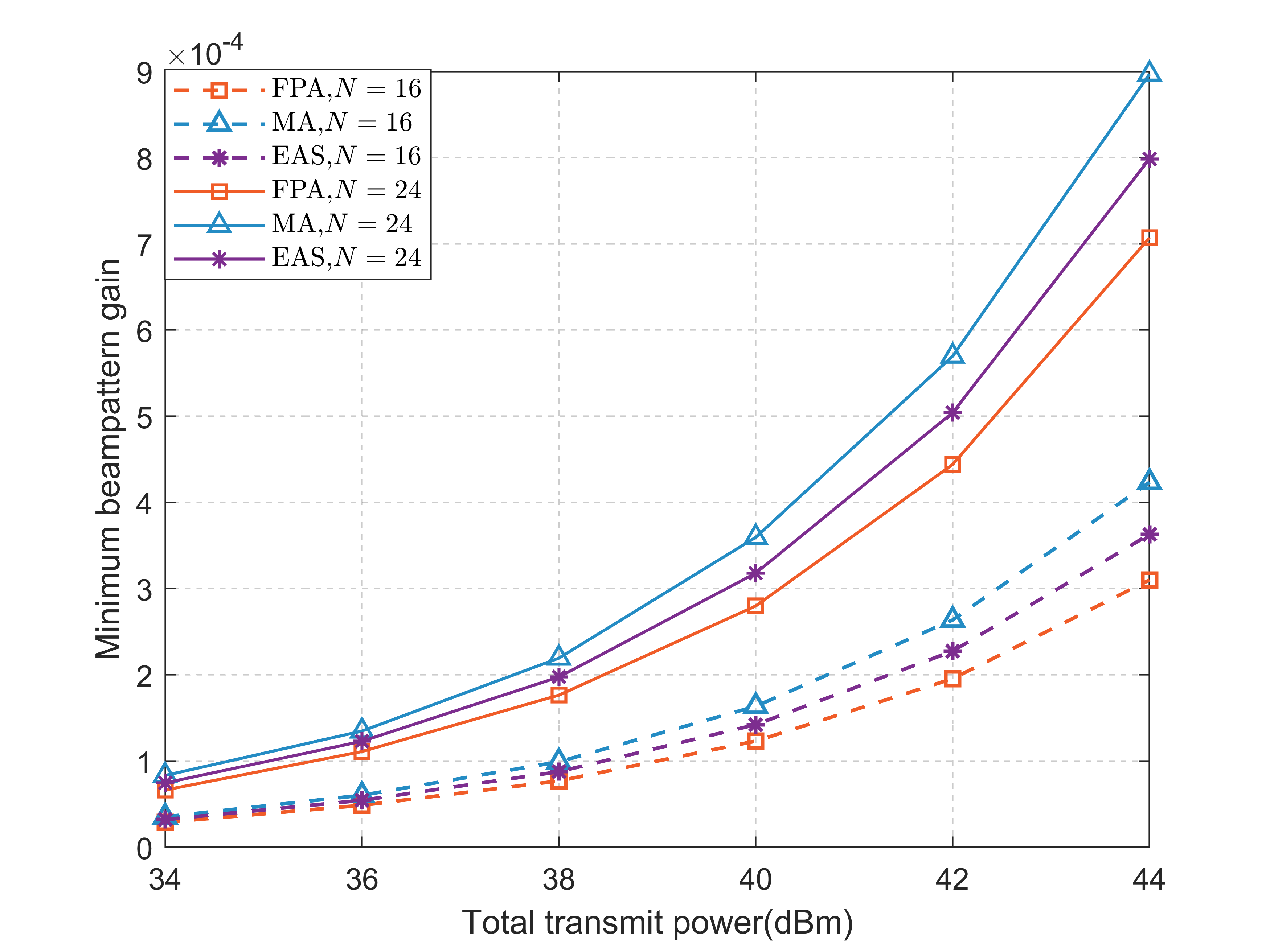}
    \caption{Minimum beampattern gain versus transmit power $P_0$.}
    \label{fig:Power}
\end{figure}

\begin{figure}
    \centering
    \includegraphics[width=1.01\linewidth]{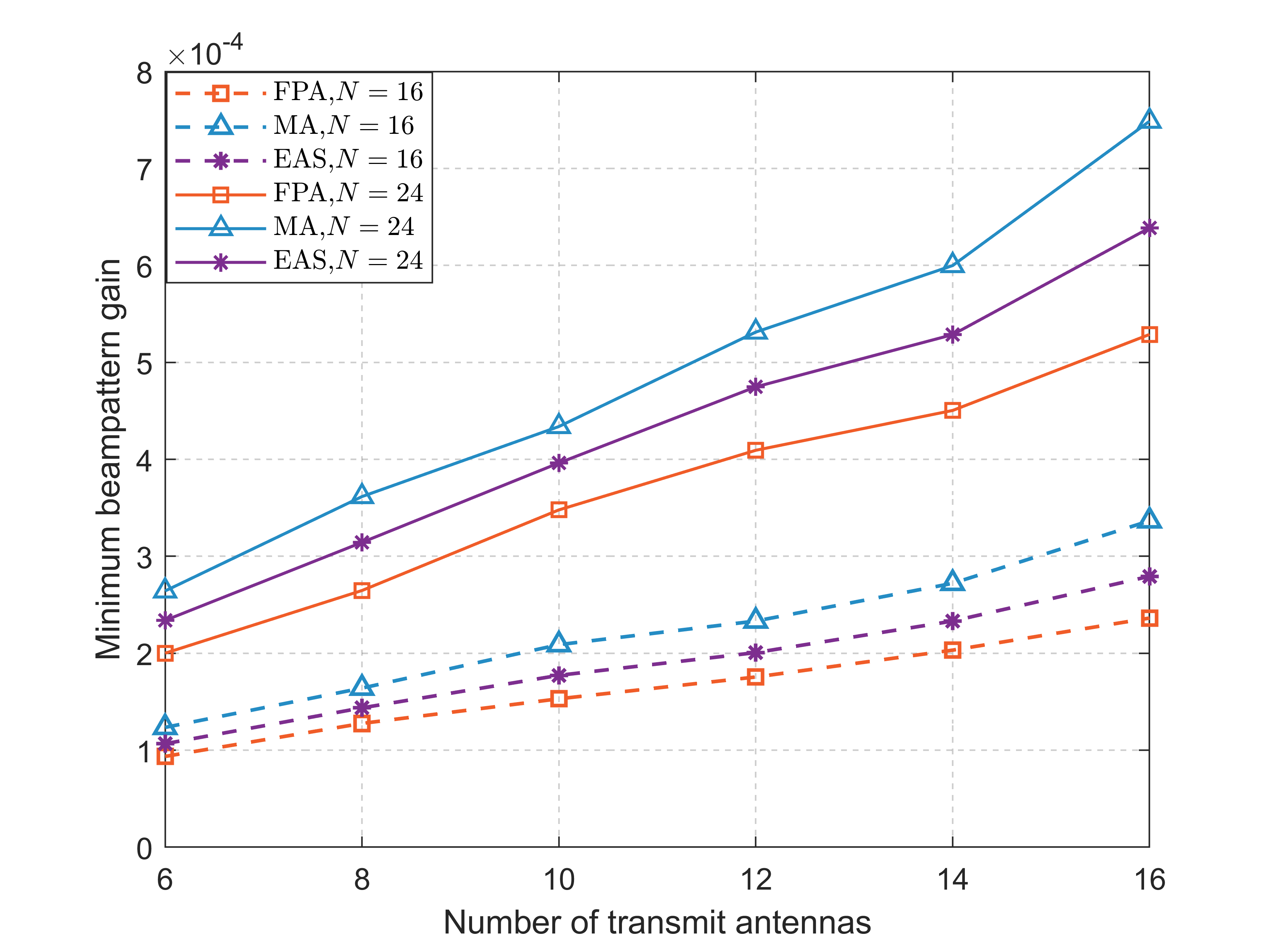}
    \caption{Minimum beampattern gain versus the number of transmit  antennas $M$.}
    \label{fig:M}
\end{figure}

Furthermore,  we present the  average channel gain of user equivalent channel $\|\mathbf{h}_{k}\|^2$ and  cross-correlation coefficient $\rho$  over iteration index in Fig. \ref{fig:User_channel}. The  cross-correlation coefficient of the equivalent user channels  $\rho$ is defined as $\frac1{K(K-1)}\sum_{1\leq k\neq q\leq K}\frac{\left|\mathbf{h}_k(\tilde{\boldsymbol{t}})^H\mathbf{h}_q(\tilde{\boldsymbol{t}})\right|}{\|\mathbf{h}_k(\tilde{\boldsymbol{t}})\|_2\|\mathbf{h}_q(\tilde{\boldsymbol{t}})\|_2}$. Interestingly, we find that during the procedure of the joint optimization, the variation of  $\|\mathbf{h}_{k}\|^2$ and $\rho$ can be divided into two phases. In the first phase, the user channel gain decreases by approximately 0.2\%, while the user channel similarity decreases by about 10.25\%. Therefore, the gain in the ISAC system performance during this phase is partly attributed to the alleviation of multiuser interference, rather than merely enhancing the signal power of individual users. In the second phase, the user channel gain increases by approximately 4.75\%, and the channel similarity increases by about 0.78\%. This can be attributed to the fact that the channel gain of the BS-RIS link is increasing along with the optimization process to enhance the sensing performance. Moreover, since the channel gain of the BS-RIS link is increasing during the second phase, it is reasonable to expect that both the gain and similarity of the user equivalent channel are also increasing during this phase.

\subsection{ Performance Comparison with Benchmark Schemes}
To comprehensively illustrate the  advantages of  MAs in enhancing the  sensing and communication performance, we propose the following two baseline schemes:
\begin{enumerate}
	\item {\bf{\text{Fixed position antenna (FPA)}}}: The BS is equipped with a UPA based on FPA configuration, with $M$ antennas spaced between interval of $\frac{\lambda}{2}$.

	\item {\bf{\text{Exhaustive antenna selection (EAS)}}}: The BS is equipped with an FPA-based UPA, consisting of $2M$ antennas spaced at $\frac{\lambda}{2}$ intervals. From these, $M$ antennas are selected through an exhaustive search.
\end{enumerate}

In Fig. \ref{fig:Power}, we illustrate the beampattern gain of the proposed and benchmark schemes versus the maximum transmit power $P_0$. Firstly, it can be clearly observed  that the ISAC system  with $N=24$ RIS elements exhibits approximately a 110\% performance improvement over the system with $N=16$ under three different transmit antenna configurations, demonstrating the importance of deploying RIS in ISAC systems for sensing dead zones. Additionally, the MA scheme, which fully utilizes spatial DoFs, achieves about 30\% performance gain over the FPA scheme in all scenarios. This is achieved by simultaneously improving the channel environment, suppressing multi-user interference, and optimizing the geometric characteristics of the antenna array.

Fig. \ref{fig:M} shows the minimum beampattern gain versus the number of transmit antennas. It illustrates that the beampattern gain increases in all schemes due to the spatial diversity gain and array gain introduced by additional antennas. Notably, when the number of antennas doubles from 6 to 12 in the  MA-aided scheme with $N = 24$, the beampattern gain increases from  from $2.73\times10^{-4}$ to $5.34\times10^{-4}$, achieving a gain of 95.6\%. This indicates that increasing the number of  MAs can significantly enhance the sensing performance by leveraging the spatial DoFs.

\begin{figure}
    \centering
    \includegraphics[width=1.05\linewidth]{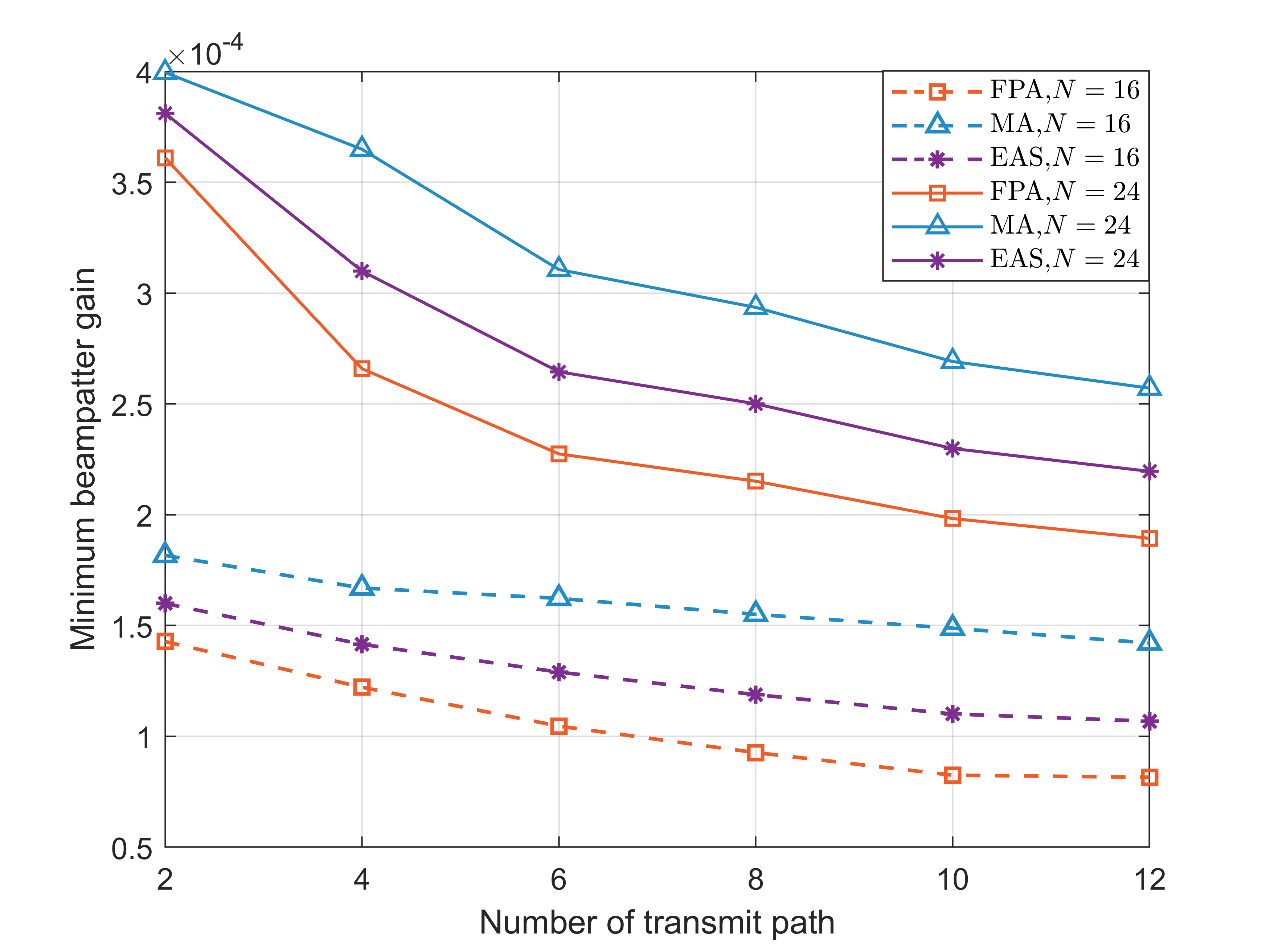}
    \caption{Minimum beampattern gain versus the number of thansmit paths $L$.}
    \label{fig:L}
\end{figure}

The relationship between the minimum beampattern gain and the number of transmit paths is demonstrated in Fig.  \ref{fig:L}. It is worth noting that the sensing performance of  MA-aided DFRC systems declines with the increase in $L$, which is contrary to the property that channel capacity tends to increase with the number of transmit paths\cite{ma2023mimo}. This phenomenon can be attributed to the negative impact of electromagnetic wave scattering on sensing capabilities. 


\section{CONCLUSION}
In this paper, we investigated an  MA and RIS-aided ISAC system, where the DFRC BS is equipped with  MAs to enhance both sensing and communication performance.  Aiming at  maximizing the minimum beampattern gain at the RIS towards the desired sensing angles, we jointly optimized the transmit beamforming at the BS, the phase shifts of  the RIS and the positions of the  MAs, subject to the  user SINR constraint and the transmit power constraint at the BS. Due to the highly non-convex nature of the resultant problem, we developed an AO-based algorithm utilizing SDR, SRCR, and SCA techniques. Specifically, the SRCR algorithm was utilized for optimizing the phase shifts at the RIS, owing to the potential nonconvergence characteristics caused by SDR.  Simulation results illustrated significant advantages of the  MA and RIS-aided system over other baseline schemes in ISAC systems. In addition, this paper meticulously discussed the the variation in channel characteristics during the procedure of antenna position optimization. It is observed that the energy of the user channels initially decreases before increasing during the optimization process of the iterative algorithm, which is owing to the deployment of the RIS.

\begin{figure*}[!ht]
\begin{equation}
        \begin{aligned}\frac{\partial \tilde{I}(\boldsymbol{t}_m)}{\partial x_m}&=-\frac{4\pi}\lambda\sum_{i=1}^{L_t^k-1}\sum_{j=i+1}^{L_t^k}\left[\widetilde{\mathbf R}_k \right]_{m,m}\left|\left[\boldsymbol{Q}_k\right]_{i,j}\right|\left(-\cos\phi_{k,i}^t\sin\theta_{k,i}^t+\cos\phi_{k,j}^t\sin\theta_{k,j}^t\right)\sin\left(\frac{2\pi}{\lambda}\left(\rho_{t,k}^i(\boldsymbol{t}_m)-\rho_{t,k}^j(\boldsymbol{t}_m)\right)+\angle\left[\boldsymbol{Q}_k\right]_{i,j}\right)\\
    &\quad-\frac{4\pi}\lambda\sum_{i=1}^{L_t-1}\sum_{j=i+1}^{L_t}\left[\widetilde{\mathbf R}_k \right]_{m,m}\left|\left[\boldsymbol{P}_k\right]_{i,j}\right|\left(-\cos\phi_{i}^t\sin\theta_{i}^t+\cos\phi_{j}^t\sin\theta_{j}^t\right)\sin\left(\frac{2\pi}{\lambda}\left(\rho_{t}^i(\boldsymbol{t}_m)-\rho_{t}^j(\boldsymbol{t}_m)\right)+\angle\left[\boldsymbol{P}_k\right]_{i,j}\right)\\
    &\quad-\frac{4\pi}\lambda\sum_{i=1}^{L_t}\sum_{j=1}^{L_t^k}\left[\widetilde{\mathbf R}_k \right]_{m,m}\left|\left[\boldsymbol{p}_k\right]_i\right|\left|\left[
    \boldsymbol{q}_k\right]_j\right|\left(-\cos\phi_{k,i}^t\sin\theta_{k,i}^t+\cos\phi_{j}^t\sin\theta_{j}^t\right)\sin\left(\frac{2\pi}{\lambda}\left(\rho_{t}^i(\boldsymbol{t}_m)-\rho_{t,k}^j(\boldsymbol{t}_m)\right)+\angle \left[\boldsymbol{q}_k\right]_j-\angle \left[\boldsymbol{p}_k\right]_i\right)\\
    &\quad-\frac{4\pi}\lambda\sum_{i=1}^{L_t}\left|
    \tilde{a}_k\right|\left|\left[ \boldsymbol{p}_k\right]_i\right|\cos\phi_{i}^t\sin\theta_{i}^t\sin\left(\frac{2\pi}\lambda\rho_{t}^i(\boldsymbol{t}_m)-\angle\left[
    \boldsymbol{p}_k\right]_i+\angle\tilde{a}_k\right)\\
    &\quad-\frac{4\pi}\lambda\sum_{j=1}^{L_t^k}\left|\tilde{a}_k\right|\left|\left[\boldsymbol{q}_k\right]_j\right|\cos\phi_{k,j}^t\sin\theta_{k,j}^t\sin\left(\frac{2\pi}\lambda\rho_{t,k}^i(\boldsymbol{t}_m)-\angle\left[\boldsymbol{q}_k\right]_j+\angle\tilde{a}_k\right)
\end{aligned}\label{eq:partial_x}
\end{equation}
\begin{equation}
    \begin{aligned}\frac{\partial \tilde{I}(\boldsymbol{t}_m)}{\partial y_m}&=-\frac{4\pi}\lambda\sum_{i=1}^{L_t^k-1}\sum_{j=i+1}^{L_t^k}\left[\widetilde{\mathbf R}_k \right]_{m,m}\left|\left[\boldsymbol{Q}_k\right]_{i,j}\right|\left(-\cos\theta_{k,i}^t+\cos\theta_{k,j}^t\right)\sin\left(\frac{2\pi}{\lambda}\left(\rho_{t,k}^i(\boldsymbol{t}_m)-\rho_{t,k}^j(\boldsymbol{t}_m)\right)+\angle\left[\boldsymbol{Q}_k\right]_{i,j}\right)\\
    &\quad-\frac{4\pi}\lambda\sum_{i=1}^{L_t-1}\sum_{j=i+1}^{L_t}\left[\widetilde{\mathbf R}_k \right]_{m,m}\left|\left[\boldsymbol{P}_k\right]_{i,j}\right|\left(-\cos\theta_{i}^t+\cos\theta_{j}^t\right)\sin\left(\frac{2\pi}{\lambda}\left(\rho_{t}^i(\boldsymbol{t}_m)-\rho_{t}^j(\boldsymbol{t}_m)\right)+\angle\left[\boldsymbol{P}_k\right]_{i,j}\right)\\
    &\quad-\frac{4\pi}\lambda\sum_{i=1}^{L_t}\sum_{j=1}^{L_t^k}\left[\widetilde{\mathbf R}_k \right]_{m,m}\left|\left[\boldsymbol{p}_k\right]_i\right|\left|\left[
    \boldsymbol{q}_k\right]_j\right|\left(-\cos\theta_{k,i}^t+\cos\theta_{j}^t\right)\sin\left(\frac{2\pi}{\lambda}\left(\rho_{t}^i(\boldsymbol{t}_m)-\rho_{t,k}^j(\boldsymbol{t}_m)\right)+\angle \left[\boldsymbol{q}_k\right]_j-\angle \left[\boldsymbol{p}_k\right]_i\right)\\
    &\quad-\frac{4\pi}\lambda\sum_{i=1}^{L_t}\left|
    \tilde{a}_k\right|\left|\left[ \boldsymbol{p}_k\right]_i\right|\cos\theta_{i}^t\sin\left(\frac{2\pi}\lambda\rho_{t}^i(\boldsymbol{t}_m)-\angle\left[
    \boldsymbol{p}_k\right]_i+\angle\tilde{a}_k\right)\\
    &\quad-\frac{4\pi}\lambda\sum_{j=1}^{L_t^k}\left|\tilde{a}_k\right|\left|\left[\boldsymbol{q}_k\right]_j\right|\cos\theta_{k,j}^t\sin\left(\frac{2\pi}\lambda\rho_{t,k}^i(\boldsymbol{t}_m)-\angle\left[\boldsymbol{q}_k\right]_j+\angle\tilde{a}_k\right)
\end{aligned}\label{eq:partial_y}
\end{equation}
\end{figure*}

\begin{appendices}
\section{}
The relative terms of  gradient vector $\nabla\tilde{I}(\boldsymbol{t}_m)=\begin{bmatrix}\frac{\partial\tilde{I}(\boldsymbol{t}_m)}{\partial x_m},\frac{\partial\tilde{I}(\boldsymbol{t}_m)}{\partial y_m}\end{bmatrix}^T$  are given in (\ref{eq:partial_x}), (\ref{eq:partial_y}). For convenience of expression, we define 
\begin{equation}
    \iota_{k,i,j}^1(\boldsymbol{t}_m)=\frac{2\pi}{\lambda}\left(\rho_{t,k}^i(\boldsymbol{t}_m)-\rho_{t,k}^j(\boldsymbol{t}_m)\right)+\angle\left[\boldsymbol{Q}_k\right]_{i,j},
\end{equation}
\begin{equation}
     \iota_{k,i,j}^2(\boldsymbol{t}_m)=\frac{2\pi}{\lambda}\left(\rho_{t}^i(\boldsymbol{t}_m)-\rho_{t}^j(\boldsymbol{t}_m)\right)+\angle\left[\boldsymbol{P}_k\right]_{i,j},
\end{equation}
\begin{equation}
     \iota_{k,i,j}^3(\boldsymbol{t}_m)=\frac{2\pi}{\lambda}\left(\rho_{t}^i(\boldsymbol{t}_m)-\rho_{t,k}^j(\boldsymbol{t}_m)\right)+\angle \left[\boldsymbol{q}_k\right]_j-\angle \left[\boldsymbol{p}_k\right]_i,
\end{equation}
\begin{equation}            \kappa_{k,i}^1(\boldsymbol{t}_m)=\frac{2\pi}\lambda\rho_{t}^i(\boldsymbol{t}_m)-\angle\left[
    \boldsymbol{p}_k\right]_i+\angle\tilde{a}_k,
\end{equation}
\begin{equation}            \kappa_{k,j}^2(\boldsymbol{t}_m)=\frac{2\pi}\lambda\rho_{t,k}^i(\boldsymbol{t}_m)-\angle\left[\boldsymbol{q}_k\right]_j+\angle\tilde{a}_k,
\end{equation}
and relative terms of the Hessian matrix $\left.\nabla^{2}\tilde{I}\left(\boldsymbol{t}_{m}\right)=\left[\begin{array}{cc}\frac{\partial^{2}\tilde{I}\left(\boldsymbol{t}_{m}\right)}{\partial x_{m}\partial x_{m}}&\frac{\partial^{2}\tilde{I}\left(\boldsymbol{t}_{m}\right)}{\partial x_{m}\partial y_{m}}\\\frac{\partial^{2}\tilde{I}\left(\boldsymbol{t}_{m}\right)}{\partial y_{m}\partial x_{m}}&\frac{\partial^{2}\tilde{I}\left(\boldsymbol{t}_{m}\right)}{\partial y_{m}\partial y_{m}}\end{array}\right.\right]$ are given in (\ref{eq:partial_y_y})-(\ref{eq:partial_x_y}). Specifically, $\frac{\partial^{2}\tilde{I}\left(\boldsymbol{t}_{m}\right)}{\partial y_{m}\partial x_{m}}=\frac{\partial^{2}\tilde{I}\left(\boldsymbol{t}_{m}\right)}{\partial x_{m}\partial y_{m}}$,  thus $\frac{\partial^{2}\tilde{I}\left(\boldsymbol{t}_{m}\right)}{\partial y_{m}\partial x_{m}}$ is omitted for simplicity.
\end{appendices}

\begin{figure*}[hb]

\begin{equation}
    \begin{aligned}\frac{\partial^2 \tilde{I}(\boldsymbol{t}_m)}{\partial y_m \partial y_m}&=-\frac{8\pi^2}\lambda\sum_{i=1}^{L_t^k-1}\sum_{j=i+1}^{L_t^k}\left[\widetilde{\mathbf R}_k \right]_{m,m}\left|\left[\boldsymbol{Q}_k\right]_{i,j}\right|\left(-\cos\theta_{k,i}^t+\cos\theta_{k,j}^t\right)^2\cos\left(\iota_{k,i,j}^1(\boldsymbol{t}_m)\right)\\
    &\quad-\frac{8\pi^2}\lambda\sum_{i=1}^{L_t-1}\sum_{j=i+1}^{L_t}\left[\widetilde{\mathbf R}_k \right]_{m,m}\left|\left[\boldsymbol{P}_k\right]_{i,j}\right|\left(-\cos\theta_{i}^t+\cos\theta_{j}^t\right)^2\cos\left(\iota_{k,i,j}^2(\boldsymbol{t}_m)\right)\\
    &\quad-\frac{8\pi^2}\lambda\sum_{i=1}^{L_t}\sum_{j=1}^{L_t^k}\left[\widetilde{\mathbf R}_k \right]_{m,m}\left|\left[\boldsymbol{p}_k\right]_i\right|\left|\left[
    \boldsymbol{q}_k\right]_j\right|\left(-\cos\theta_{k,i}^t+\cos\theta_{j}^t\right)^2\cos\left(\iota_{k,i,j}^3(\boldsymbol{t}_m)\right)\\
    &\quad-\frac{8\pi^2}\lambda\sum_{i=1}^{L_t}\left|
    \tilde{a}_k\right|\left|\left[ \boldsymbol{p}_k\right]_i\right|\cos^2\theta_{i}^t\cos\left(\kappa_{k,i}^1(\boldsymbol{t}_m)\right)-\frac{8\pi^2}\lambda\sum_{j=1}^{L_t^k}\left|\tilde{a}_k\right|\left|\left[\boldsymbol{q}_k\right]_j\right|\cos^2\theta_{k,j}^t\cos\left(\kappa_{k,j}^2(\boldsymbol{t}_m)\right)
\end{aligned}\label{eq:partial_y_y}
\end{equation}

\begin{equation}
    \begin{aligned}\frac{\partial^2 \tilde{I}(\boldsymbol{t}_m)}{\partial x_m \partial x_m}&=-\frac{8\pi^2}\lambda\sum_{i=1}^{L_t^k-1}\sum_{j=i+1}^{L_t^k}\left[\widetilde{\mathbf R}_k \right]_{m,m}\left|\left[\boldsymbol{Q}_k\right]_{i,j}\right|\left(-\cos\phi_{k,i}^t\sin\theta_{k,i}^t+\cos\phi_{k,j}^t\sin\theta_{k,j}^t\right)^2\cos\left(\iota_{k,i,j}^1(\boldsymbol{t}_m)\right)\\
    &\quad-\frac{8\pi^2}\lambda\sum_{i=1}^{L_t-1}\sum_{j=i+1}^{L_t}\left[\widetilde{\mathbf R}_k \right]_{m,m}\left|\left[\boldsymbol{P}_k\right]_{i,j}\right|\left(-\cos\phi_{i}^t\sin\theta_{i}^t+\cos\phi_{j}^t\sin\theta_{j}^t\right)^2\cos\left(\iota_{k,i,j}^2(\boldsymbol{t}_m)\right)\\
    &\quad-\frac{8\pi^2}\lambda\sum_{i=1}^{L_t}\sum_{j=1}^{L_t^k}\left[\widetilde{\mathbf R}_k \right]_{m,m}\left|\left[\boldsymbol{p}_k\right]_i\right|\left|\left[
    \boldsymbol{q}_k\right]_j\right|\left(-\cos\phi_{k,i}^t\sin\theta_{k,i}^t+\cos\phi_{j}^t\sin\theta_{j}^t\right)^2\cos\left(\iota_{k,i,j}^3(\boldsymbol{t}_m)\right)\\
    &\quad-\frac{8\pi^2}\lambda\sum_{i=1}^{L_t}\left|
    \tilde{a}_k\right|\left|\left[ \boldsymbol{p}_k\right]_i\right|\cos^2\phi_{i}^t\sin^2\theta_{i}^t\cos\left(\kappa_{k,i}^1(\boldsymbol{t}_m)\right)-\frac{8\pi^2}\lambda\sum_{j=1}^{L_t^k}\left|\tilde{a}_k\right|\left|\left[\boldsymbol{q}_k\right]_j\right|\cos^2\phi_{k,j}^t\sin^2\theta_{k,j}^t\cos\left(\kappa_{k,j}^2(\boldsymbol{t}_m)\right)\label{eq:partial_x_x}
\end{aligned}
\end{equation}

\begin{equation}
      \begin{aligned}\frac{\partial^2 \tilde{I}(\boldsymbol{t}_m)}{\partial x_m \partial y_m}&=-\frac{8\pi^2}\lambda\sum_{i=1}^{L_t^k-1}\sum_{j=i+1}^{L_t^k}\left[\widetilde{\mathbf R}_k \right]_{m,m}\left|\left[\boldsymbol{Q}_k\right]_{i,j}\right|\left(-\cos\phi_{k,i}^t\sin\theta_{k,i}^t+\cos\phi_{k,j}^t\sin\theta_{k,j}^t\right)\left(-\cos\theta_{k,i}^t+\cos\theta_{k,j}^t\right)\cos\left(\iota_{k,i,j}^1(\boldsymbol{t}_m)\right)\\
    &\quad-\frac{8\pi^2}\lambda\sum_{i=1}^{L_t-1}\sum_{j=i+1}^{L_t}\left[\widetilde{\mathbf R}_k \right]_{m,m}\left|\left[\boldsymbol{P}_k\right]_{i,j}\right|\left(-\cos\phi_{i}^t\sin\theta_{i}^t+\cos\phi_{j}^t\sin\theta_{j}^t\right)\left(-\cos\theta_{i}^t+\cos\theta_{j}^t\right)\cos\left(\iota_{k,i,j}^2(\boldsymbol{t}_m)\right)\\
    &\quad-\frac{8\pi^2}\lambda\sum_{i=1}^{L_t}\sum_{j=1}^{L_t^k}\left[\widetilde{\mathbf R}_k \right]_{m,m}\left|\left[\boldsymbol{p}_k\right]_i\right|\left|\left[
    \boldsymbol{q}_k\right]_j\right|\left(-\cos\phi_{k,i}^t\sin\theta_{k,i}^t+\cos\phi_{j}^t\sin\theta_{j}^t\right)\left(-\cos\theta_{k,i}^t+\cos\theta_{j}^t\right)\cos\left(\iota_{k,i,j}^3(\boldsymbol{t}_m)\right)\\
    &\quad-\frac{8\pi^2}\lambda\sum_{i=1}^{L_t}\left|
    \tilde{a}_k\right|\left|\left[ \boldsymbol{p}_k\right]_i\right|\cos\phi_{i}^t\sin\theta_{i}^t\cos\theta_{i}^t\cos\left(\kappa_{k,i}^1(\boldsymbol{t}_m)\right)-\frac{8\pi^2}\lambda\sum_{j=1}^{L_t^k}\left|\tilde{a}_k\right|\left|\left[\boldsymbol{q}_k\right]_j\right|\cos\phi_{k,j}^t\sin\theta_{k,j}^t\cos\theta_{k,j}^t\cos\left(\kappa_{k,j}^2(\boldsymbol{t}_m)\right)
\end{aligned}\label{eq:partial_x_y}
\end{equation}   
\end{figure*}

\bibliography{ref.bib}
\bibliographystyle{ieeetr}

\vfill

\end{document}